\documentclass[10pt,twocolumn,twoside] {IEEEtran}

\usepackage{cite}
\usepackage{graphicx}
\usepackage{booktabs}
\usepackage{amsmath}
\usepackage{amssymb} 
\usepackage{midfloat}
\usepackage{tabularx}
\usepackage{bm}
\usepackage{extarrows}
\usepackage{subcaption}
\usepackage{color, cite}
\usepackage{threeparttable}
\usepackage{algpseudocode}
\usepackage{algorithmicx}
\usepackage{dcolumn}
\usepackage{lipsum} 
\usepackage{mathtools}
\usepackage{multirow}

\usepackage[ruled, linesnumbered]{algorithm2e}

\newcommand{\myindent}[2]{
	\newline\makebox[#1cm]{}}

\graphicspath{{figures/}}

\usepackage[table,xcdraw]{xcolor}
\usepackage{easyReview}
\begin{document}
	
\title{Identification of Ghost Targets for Automotive Radar in the Presence of Multipath }
\author{Le Zheng,  \emph{Senior Member}, \emph{IEEE}, Jiamin Long,  Marco Lops,  \emph{ Fellow}, \emph{IEEE}\\
	Fan Liu,  \emph{Senior Member}, \emph{IEEE}, Xueyao Hu
	\thanks{
		This research was funded by  the National Key R$\&$D Program of China (Grant No. 2018YFE0202101, 2018YFE0202102, 2018YFE0202103);
		
		Le Zheng, Jiamin Long and Xueyao Hu are with the Radar Research Laboratory, School of Information
		and Electronics, Beijing Institute of Technology, Beijing 100081, China.
		Le Zheng  and Xueyao Hu are also with the Chongqing Innovation	Center, Beijing Institute of Technology, Chongqing 401120, China(e-mail:le.zheng.cn@gmail.com, jiaminlong0548@163.com, xueyao.hu@qq.com).
		
		Marco Lops is with the Department of Electrical and Information Technology (DIETI), University of Naples Federico II, 80138 Naples, Italy, and also with the Consorzio Nazionale Interuniversitario per le Telecomunicazioni, 43124 Parma, Italy (e-mail: lops@unina.it).
		
		Fan Liu is with the Department of Electronic and Electrical Engineering, Southern University of Science and Technology, Shenzhen 518055, China (e-mail: liuf6@sustech.edu.cn).
	}
}
\maketitle
\begin{abstract}
	Colocated multiple-input multiple-output (MIMO) technology has been widely used in automotive radars as it provides accurate angular estimation of the objects with relatively small number of transmitting and receiving antennas.
	 Since the Direction Of Departure (DOD) and the Direction Of Arrival (DOA) of {\em line-of-sight} targets coincide, MIMO signal processing allows forming a larger virtual array for angle finding.  However,  multiple paths  impinging the receiver is a major limiting factor, in that radar signals may bounce off obstacles, creating echoes for which the DOD does not equal the DOA. Thus, in complex scenarios with multiple scatterers, the direct paths of the intended targets may be corrupted by indirect paths from other objects, which leads to inaccurate angle estimation or ghost targets. 
	In this paper, we focus on
	detecting the presence of ghosts due to multipath by regarding it as the problem of deciding between a {\em composite} hypothesis,  ${\cal H}_0$ say, that the observations only contain an unknown number of direct paths sharing the same (unknown) DOD's and DOA's, and a {\em composite} alternative, ${\cal H}_1$ say, that the observations also contain an unknown number of indirect paths, for which DOD's and DOA's do not coincide.
	We exploit the Generalized Likelihood Ratio Test (GLRT) philosophy to determine the detector structure, wherein the unknown parameters are replaced by carefully designed estimators. The angles of both the active direct paths and of the multi-paths are indeed estimated through a  sparsity-enforced Compressed Sensing (CS) approach with Levenberg-Marquardt (LM) optimization to estimate the angular parameters in the continuous domain. An extensive performance analysis is finally offered in order to validate the proposed solution.
\color{black}
\end{abstract}
\begin{IEEEkeywords}
	 Automotive radar, Colocated multiple-input multiple-output (MIMO), multipath, GLRT, group sparse. 
\end{IEEEkeywords}
\IEEEpeerreviewmaketitle
\section{Introduction}
In recent years, the need for safer  driving has led to a significant   demand for automotive radar\cite{patole_automotive_2017,engels_advances_2017,roos_radar_2019,waldschmidt_automotive_2021}. Colocated MIMO technology   has been proven to be effective in providing accurate angular estimation of objects with a relatively small number of antennas, making it popular in the automotive industry \cite{li_mimo_2007,sun_mimo_2020,hu_multi_carrier_frequency_2019}. 

One major challenge of colocated MIMO systems is the multipath reflection, where the target's echo takes multiple paths to reach the receiver, including direct and indirect paths \cite{kamann_multipath_2018, visentin_analysis_nodate, sabet_hybrid_2020, longman_multipath_2021}.
Direct paths involve the signal being transmitted from the radar to the target and then reflected back to the radar directly, while the indirect paths could bounce multiple times between reflectors. 
 Usually, due to different propagation delays,  range gating can rid of the indirect paths from the target we are trying to detect.  However, the DOD of the signal does not equal the DOA for some indirect paths,\cite{li_mimo_2007,Levy_MCRB_2023}, so the assumption of colocated MIMO does not hold. As a consequence, in multi-target scenarios the direct paths of intended targets may be corrupted by indirect paths from  other objects and applying classical angle finding algorithms may result into  degraded angle  estimation  accuracy and detection of ghost targets.

To identify ghost targets, some   researchers exploit the geometrical relationships of the {detections} in the  delay-Doppler domain. Specifically, R. Feng et al. employed  the Hough transform to explore the linear relationship of the multipath  {detections} \cite{feng_multipath_2022}. F. Ross et al.  identified the ghost targets by analyzing the Doppler distribution of moving targets\cite{Roos_dopDis_2017}. {These methods can be effective when the speed of the ghost target is significant, and the efficient utilization of Doppler information can aid in extracting geometric information from multipaths for identification. However, in  situations with densely distributed objects,  ghost targets with low speeds may couple   with the stationary objects, making it difficult to use Doppler information to identify them.}

{Several strategies for multipath ghost suppression in the} angular domain have been proposed so far, ranging from antenna design\cite{visentin_analysis_nodate,li_fans_shaped_2021} to synthetic aperture radar (SAR)\cite{manzoni_multipath_2023} and  deep learning \cite{jia_multipath_2019}.  Considering the potential advantages of indirect paths in non-line-of-sight detection or reconfigurable intelligent surface (RIS) applications\cite{wu_NLOS_2023,buzzi_foundations_2022,zhang_beampattern_2022}, accurately detecting and estimating the parameters of each path is  more valuable than simply suppressing multipath:  {this is the idea} underlying \cite{engels_automotive_2017}, where the presence of multipath reflections is detected through a Generalized Likelihood Ratio Test (GLRT).  {The detector was developed under a specific signal model where only two TX antennas are used and all indirect paths for a target are confined to a single delay-Doppler cell. However, MIMO arrays commonly used in automotive radars usually have more TX antennas\cite{li_ultra_fast_2022}. Furthermore, a single delay-Doppler cell might encompass direct and indirect paths from different targets. Given the potential model mismatch, the performance of angle estimation in \cite{engels_automotive_2017} degrades in such situations and the GLRT would fail.}

Accurate estimation of target information is crucial for ghost identification in  angular domain. In the field of bistatic MIMO radar, the angle finding methods for situations with different DOA and DOD have  been widely studied. Subspace methods, such as the two-dimensional multiple signal classification (2D-MUSIC) \cite{xiaofei_DOD_2010} and unitary-estimation of signal parameters via rotational invariance technique (U-ESPRIT) \cite{Jiang_Joint_2019} have been proposed. These methods have limitations related to the signal and noise characteristics, array geometry and computational complexity, which make them unsuitable for automotive radars.
In \cite{li_multipath_2022}, an iterative adaptive approach (IAA) based method  was employed to estimate the multi-paths for automotive radar. More recent techniques based on compressed sensing (CS) theory
\cite{zheng_does_2017} provide  an alternative for jointly estimating the DOD and DOA \cite{xie_sparsity_2019,Wen_2D_DOD_2023}. The performance of these methods depends on the designed dictionaries and gridding scheme in  angular domain.
  However, as the paths are usually specified by parameters in   continuous domain, the discretization usually leads to model mismatch and degradation in estimation \cite{Jokanovic_Reduced_2015,zheng_super_resolution_2017}.

In this paper, we further investigate ghost targets identification in the angular domain, to the end of detecting the indirect paths and allowing their removal, so as to preserve only the direct paths from the target. Two types of paths are considered in our analysis: direct paths, exhibiting the same DOD and DOA and first-order paths (more on this in Section II) whose DOD does not equal DOA. After deriving the MIMO radar signal model, the problem of first-order paths existence and identification is stated as a binary decision problem between a composite hypothesis, ${\cal H}_0$ say, that the observations only contain an unknown number of direct paths sharing the same (unknown) DOD's and DOA's, and a {\em composite} alternative, ${\cal H}_1$ say, that the observations also contain an unknown number of indirect paths, for which DOD's and DOA's do not coincide. In this context, we resort to the GLRT philosophy to determine the detector structure, wherein the unknown parameters are replaced by carefully designed estimators. 
In particular, to estimate the angle of the paths under the two alternative hypotheses, we develop CS  methods in continuous domain for the cases with and without first-order paths, respectively. Specifically, in the situation with first-order paths, the algorithm is designed with group-sparsity enforced structure to take advantage of the reversibility of propagation path. To improve the convergence performance, we adopt a Levenberg-Marquardt (LM) optimization approach for to accelerate the execution of the algorithm. The proposed  method has shown the superior performance  over existing methods by simulation. An extensive performance assessment is finally offered in order to validate the proposed strategy.

The remainder of the paper is organized as follows: In Section II, we present the   signal model of multipath reflection. Section III details the proposed detector and derived its exact theoretical performance. In Section IV, we describe the proposed angle estimation methods under different situations.  In Section V, we present the simulation results, and finally, Section VI concludes the paper.

{\em{Notation} }: 
The transposition, Hermitian transposition, inversion, pseudo-inversion, Kronecker product, Khatri-Rao (KR) product, Hadamard product and direct sum  operations are denoted by   $(\cdot)^T$, $(\cdot)^H$, $(\cdot)^{-1}$,  $(\cdot)^{\dagger}$, $\otimes$ $\circ$, $\odot$, $\oplus$, respectively. Matrix $\mathbf{X}$ and vector $\mathbf{x}$  are indicated in boldface. The notation $\rm{diag}(\mathbf{X})$ denotes   the operation of extracting elements from the diagonal of  $\mathbf{X}$ to form a new vector.  $\|\mathbf{x}\|_2=\sqrt{\sum_i x_i^2}$ denotes the $\ell_2$-norm,  $\|\mathbf{x}\|_1=\sum\left|x_i\right|$ denotes the $\ell_1$-norm. 
 $\mathcal{R}(\mathbf{X})$ denote the range-span of the matrix $\mathbf{X}$. $\mathbf{x}^{(k)}$ denotes the value of $\mathbf{x}$ at the $k$-th iteration and $\mathbf{x}^{(k,j)}$ denotes the value of $\mathbf{x}^{(k)}$ at the $j$-th iteration.  $ \bm I_{n}$ being the  ${n}\times{n}$ identity matrix. For $\mathbf{X}$ , the $n$-th column vector and $(m, n)$-th element are denoted by $\mathbf{X}(n)$ and $[\mathbf{X}]_{m,n}$, respectively, while the $m$-th element of vector $\mathbf{x}$  is given by $[\mathbf{x}]_m$ .

\section{Signal Model and Problem Formulation}
State-of-the-art automotive radars usually employ Frequency Modulated Continuous Wave (FMCW) sequences to enable high-resolution estimation of target range and velocity \cite{Rohling_984612,kronauge_new_2014}, and adopt colocated MIMO technology to synthesize a large virtual array for accurate angle estimation using multiple transmit and receive antennas. Consider a colocated MIMO radar system with $M_T$ transmit antennas emitting
orthogonal FMCW sequences\cite{doris_reframing_2022} and $M_R$ receive antennas. 
At the receiver end, the signal at each antenna  undergoes the usual processing to extract the contribution of each transmit antenna and synthesize a MIMO channel with $M_TM_R$ elements. 
This signal is then processed via fast Fourier transform  (FFT) along fast and slow time to obtain the delay-Doppler  profile of the echo path\cite{patole_automotive_2017}.  Finally, the virtual array response of the detected target can be constructed to estimate the direction of targets\cite{sun_mimo_2020}.  

The multipath scenario can be visualized as a radar emitting signals that bounce off a target and a reflector. 
As   depicted in Fig.  \ref{fig:indirect_geometry}, where the target  is placed at position A and the reflector is located at point B, the signals received by the radar can take   different paths as follows:
\begin{figure}
	\centering
	\includegraphics{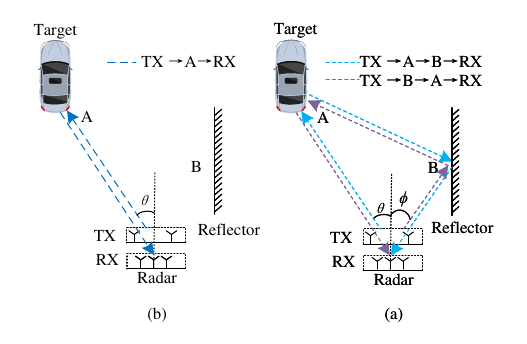}
	\caption{   (a) A direct path, (b) A pair of first-order paths.}
	\label{fig:indirect_geometry}
\end{figure} 
\begin{itemize}
	\item \textit{Direct path} : The  shortest path between the radar and the target, where the departure and arrival angles of the direct path are equal to the target angle as shown in Fig.  \ref{fig:indirect_geometry}a.
	\item \textit{First-order paths} : The indirect paths involve a single bounce at the reflector on the way of departure or arrival, resulting in a longer delay compared to the direct path. As  shown in Fig.  \ref{fig:indirect_geometry} b, the DOD's of the first-order paths do not equal the respective DOA's. 
	\item \textit{Higher-order paths} :  The indirect paths involve  more bounces before the echo reaches
	the receiver. However, due to the attenuation caused by scattering at the target and reflectors, higher-order
	paths are normally weak, and may thus be neglected \cite{wang_multipath_2022,feng_multipath_2022}.
\end{itemize}

In automotive radar, delay and Doppler information of direct path yields target range and velocity\cite{engels_advances_2017}, respectively. Both DOD and  DOA equal the angle of far-field targets, enabling the virtual array to form an aperture larger than the physical aperture of the radar, thereby enhancing angular resolution and accuracy of angle estimation\cite{li_mimo_2007,baral_joint_2021,li_ultra_fast_2022}. Although Fig. 1 depicts, for simplicity reasons, a single target scenario, the situation we consider here is one wherein multiple reflecting objects are present in the radar field of view, whereby the direct paths generated by the intended targets may end up with being corrupted by the first-order paths generated by other reflecting objects. 

In the case of multiple targets, the array response in a detected delay-Doppler cell   may include multiple types of paths originating from different target-reflector pairs. We then develop a general signal model in a detected delay-Doppler cell that considers  $K_0$ direct paths and $K_1$ pairs of first-order  paths. The general model of the virtual MIMO array signal $\bm z \in \mathbb{C}^{M_TM_R\times 1}$ can be expressed as follows
\begin{eqnarray}
	\label{eq:z_theta}
	\bm z &	=&\sum_{k=1}^{{K_1}}        \left(  {\beta}_{k,1} 
	\mathbf{a}_T({\vartheta}_k)\otimes \mathbf{a}_R({\varphi}_k) +{\beta}_{k,2}\mathbf{a}_T({\varphi}_k)\otimes \mathbf{a}_R({\vartheta}_k)\right) \nonumber\\
	&&+{\sum_{k=1}^{{K_0}}{\alpha}_k \mathbf{a}_T({\theta}_k)\otimes \mathbf{a}_R({\theta}_k)}  + \mathbf{w},
\end{eqnarray}
where 
\begin{itemize}
	
	\item  $\alpha_k $, ${\beta}_{k,1}  $ and ${\beta}_{k,2}$  represents the complex amplitude of the $k$-th direct path for $ k = 1, 2, \dots ,K_0$ and the $k$-th   pair   first-order  paths   for $ k = 1, 2, \dots ,K_1$ respectively. The amplitudes depend on a number of factors such as the transmit power, antenna gain pattern, path loss propagation, reflection coefficient, and matched-filter gain; 
	\item $\theta_k$  denotes the DOD of the $k$-th  direct path, which is equal to the DOA;     ${\vartheta}_k$  and $ {\varphi _k}$ denote  the DOD and DOA of  the $k$-th pair first-order  path with ${\vartheta}_k\neq {\varphi}_k$;
	\item 
	 $\mathbf{a}_T(\cdot) \in \mathbb{C}^{M_T\times1}$ and $\mathbf{a}_R(\cdot) \in \mathbb{C}^{M_R\times1}$ are the steering vectors
	\begin{eqnarray}
		\label{eq:at}
		\mathbf{a}_T(\theta) 	= \frac{1}{\sqrt{M_T}} \left[   e^{j2\pi d_{T,1} \sin(\theta)/\lambda}, e^{j2\pi d_{T,2} \sin(\theta)/\lambda}, ...\right.\nonumber\\ 
		 \left.    \myindent{2.4}, , e^{j2\pi d_{T,M_T}\sin(\theta)/\lambda} \right]^T, \\
		\label{eq:ar}
		\mathbf{a}_R(\phi)=  \frac{1}{\sqrt{M_R}} \left[ e^{j2\pi d_{R,1}\sin(\phi)/\lambda}, e^{j2\pi d_{R,2} \sin(\phi)/\lambda}, ... \right.\nonumber\\  \left.    \myindent{2.4},, e^{j2\pi  d_{R,M_R}\sin(\phi)/\lambda} \right]^T,
	\end{eqnarray}
	with $\theta$ and $\phi$ denoting the angles of $\mathbf{a}_T(\cdot)$ and $\mathbf{a}_R(\cdot)$, respectively, $\lambda$ the wavelength,
	$d_{T,m}$ and $d_{R,n}$ the relative distances of the $m$-th TX element and the $n$-th RX element from the reference array element. 
	\item $\mathbf{w}  \in \mathbb{C}^{M_TM_R\times 1}$ is the additive noise  contributed by the receiver noise and decoding residual of MIMO waveform\cite{Sun_Analysis_2014,bialer_code_2021}. In what follows, we assume $\mathbf{ w}\sim {\cal CN} (\mathbf{0}, \sigma^2\bm I _{M_TM_R})$, with $\sigma^2 $ the noise variance \cite{wang_slowtime_2020}. 
\end{itemize}

Define now ${\bm{\Theta}}_0 = [{{\theta}}_1, {{\theta}}_2,\dots, {{\theta}}_{K_0} ]^T\in \mathbb{R}^{K_0\times1}$ as the vector containing the angles of the $K_0$ direct paths. The corresponding steering matrix  is denoted as ${\mathbf{A}}({\bm{\Theta}}_0) = [{\mathbf{a}}({\theta}_1),{\mathbf{a}}({\theta}_2),\dots,{\mathbf{a}}({\theta}_{K_0})]$ $\in \mathbb{C}^{ M_TM_R \times K_0}$ where ${ \mathbf{a}}( \cdot)   = \mathbf{a}_T(\cdot)\otimes \mathbf{a}_R(\cdot)  $. In the absence of first-order paths ($K_1 = 0$), the signal model in \eqref{eq:z_theta} simplifies to
\begin{eqnarray}
	\label{eq:z_ma_1}
	\bm z = {\mathbf{A}}({\bm{\Theta}}_0) {\bm{\alpha}} + \mathbf{w},
\end{eqnarray}
where  ${\bm{\alpha}} =[\alpha_1,\alpha_2,\dots,\alpha_{K_0} ]^T\in \mathbb{C}^{K_0\times 1}$  is  the amplitude vector of direct paths.


In the presence of first-order  paths ($K_1\neq 0$), we define the DOD angle vector $\bm{\Theta}_1 = [{\vartheta}_{1},{\vartheta}_{2},\dots,{\vartheta}_{K_1}]^T \in \mathbb{R}^{K_1\times1}$, the DOA angle vector $\bm{\Phi}_1 =[{\varphi}_{1},{\varphi}_{2},\dots,{\varphi}_{K_1}]^T \in \mathbb{R}^{K_1\times1}$ and the amplitude vector  $\bm{\beta}_1 = [\beta_{{1},1},\beta_{{2},1},\dots,\beta_{{K_1},1},\beta_{{1},2},\beta_{{2},2},\dots,\beta_{{K_1},2}]^T\in \mathbb{C}^{2K_1\times1}$  for the $K_1$ pair of first-order  paths. 
Moreover, we define $ {{\bm{\Theta}}}= [{\bm{\Theta}}_1^T,{\bm{\Phi}}_1^T, \bm{\Theta}_0^T]^T\in \mathbb{R}^{(2K_1+K_0 )\times 1}$, $ {\bm{\Phi}} = [{\bm{\Phi}}_1^T,{\bm{\Theta}}_1^T, \bm{\Theta}_0^T]^T\in \mathbb{R}^{(2K_1+K_0 )\times 1}$. Denoting $\mathbf{A}_T$ and $\mathbf{A}_R$ the steering matrices of the radar TX and RX arrays, respectively, we have
\begin{eqnarray}
	\mathbf{A}_T( {\bm{\Theta}})=  
	\left[ {\mathbf{a}}_T( {\vartheta}_{1}),\dots,{\mathbf{a}}_T( {\vartheta}_{K_1}),{\mathbf{a}}_T( {\varphi}_{1}),\dots,{\mathbf{a}}_T( {\varphi}_{K_1}),\right.\nonumber\\ 
	\left.  {\mathbf{a}}_T( {\theta}_{1}),\dots,{\mathbf{a}}_T( {\theta}_{K_0}) \right],\nonumber\\ 
	\mathbf{A}_R({\bm{\Phi}} )=  
	\left[ {\mathbf{a}}_R( {\varphi}_{1}),\dots,{\mathbf{a}}_R( {\varphi}_{K_1}), {\mathbf{a}}_R( {\vartheta}_{1}),\dots,{\mathbf{a}}_R( {\vartheta}_{K_1}),\right.\nonumber\\  
	\left.  {\mathbf{a}}_R( {\theta}_{1}),\dots,{\mathbf{a}}_R( {\theta}_{K_0})   \right],\nonumber
\end{eqnarray}
and  the   signal model \eqref{eq:z_theta} can be rewritten as 
\begin{eqnarray}
	\label{eq:z_ma_2}
	\bm z =  {\mathbf{A}}({\bm{\Theta}} , {\bm{\Phi}}) {\bm{\beta}}+ \mathbf{w},
\end{eqnarray}
In the previous equation ${\mathbf{A}}({\bm{\Theta}} , {\bm{\Phi}})   =  {\mathbf{A}}_T({\bm{\Theta}}) \circ \mathbf{A}_R({\bm{\Phi}})$ denotes the response matrix, $ {\bm{\beta}} = [ {\bm{\beta}}_1^T, {\bm{\alpha}}^T]^T\in \mathbb{C}^{(2K_1+K_0)\times 1}$ is the complex amplitude vector.  Note that a pair of  first-order  paths shares same sparse pattern which is usually
smaller than the number of array elements\cite{engels_automotive_2017}, resulting in a group- sparse structure that can be employed for multipath estimation purpose.
\section{  Detection of multipath}
In the general setup outlined in the previous section, ghost identification amounts to solving a coupled detection-estimation problem, wherein we have to discriminate between a composite hypothesis,  ${\cal H}_0$ say, that the observations only contain a {\em unknown} number $K_0$ of direct paths coming from as many {\em unknown} different directions, against a {\em composite} alternative, ${\cal H}_1$ say, that the observations also contain a {\em unknown} number $K_1$ of first-order paths each  characterized by an unknown pair of angles.
In what follows, we (suboptimally) break up this problem into a two-step procedure: first, we introduce and discuss a Generalized Likelihood Ratio Test (GLRT) assuming that the number of the direct and first-order paths, as well as the corresponding angular information - i.e., the matrices ${\mathbf{A}}({\bm{\Theta}}_0)$ of \eqref{eq:z_ma_1} and ${\mathbf{A}}({\bm{\Theta}} , {\bm{\Phi}})$ of \eqref{eq:z_ma_2} - are {\em known}. Subsequently, we illustrate a number of possible techniques to provide the detector with the required information (i.e., we make it {\em implementable}), by formulating the problem of preliminary estimating these matrices as a sparse recovery problem taking full advantage of the models introduced in the previous section.
\subsection{ GLRT detector}
Assume at first that the two matrices in \eqref{eq:z_ma_1} and \eqref{eq:z_ma_2} are known, whereby we have to solve  the {\em composite} binary hypothesis test
\color{black}
\begin{eqnarray}
	\left\{\begin{array}{l}
		{\cal H}_0: \bm z =  {\mathbf{A}}(\bm{\Theta}_0)\bm{\alpha} +\mathbf{w}, \\
		{\cal H}_1: \bm z =\mathbf{A}( {\bm{\Theta}}, {\bm{\Phi}}){\bm{\beta}} +\mathbf{w},
	\end{array}\right.
\end{eqnarray}
where $\bm \alpha \in \mathbb{C}^{K_0\times 1}$ and $  \bm \beta \in \mathbb{C}^{(K_0+2K_1) \times 1}$ are {\em unknown}  parameters. Before proceeding, it is worth commenting on some constraints we want to force upon the solution of the above test, i.e.:
\begin{enumerate}
	\item  We want the test to be Constant False Alarm Rate (CFAR), i.e. its test statistic pdf under ${\cal H}_0$ {\em and} its detection threshold to be {\em functionally independent} of the noise floor {\em and} of the directions and intensities of the direct paths;
	\item  We want the resulting test to have some form of {\em optimality}, so as to use its performance as a yardstick to compare our implementable solutions to.
\end{enumerate}
Notice that, by construction, $\mathbf{A}({\bm{\Theta}},{\bm{\Phi}})$ is the matrix concatenation \cite{Scharf}
\begin{equation}
	\mathbf{A}( {\bm{\Theta}},\bm{\Phi})=[\bm E \; {\mathbf{A}}(\bm{\Theta}_0)  ],
\end{equation}
where $\bm E  = [\mathbf{A}({\bm{\Theta}}_1,\bm{\Phi}_1),\mathbf{A}({\bm{\Phi}}_1,\bm{\Theta}_1)]\in \mathbb{C}^{M_TM_R\times 2K_1}$ only depends on the DOD's and the DOA's of the first-order paths. Under the CFAR constraint outlined above, we are thus in the situation of detecting a subspace signal in subspace interference and noise of unknown level \cite[Section VIII]{Scharf}, whereby the GLRT reads
\begin{eqnarray}
	\label{eq:GLRT_un}
	\mathcal{T}_{GLRT}&=& 
	\frac{\left\|{\bm{P}}(\bm{\Theta}_0) \bm z\right\|^2}{\|\bm{P}(\bm{\Theta}, \bm{\Phi}) \bm z\|^2} \underset{{\cal H}_0}{\stackrel{{\cal H}_1}{\gtrless}} \lambda_G,
\end{eqnarray}
where ${\bm{P}}(\bm{\Theta}_0) = \bm I _{M_TM_R}- {\mathbf{A}} (\bm{\Theta}_0){\mathbf{A}}^\dagger(\bm{\Theta}_0)=
\bm P_0$ is the orthogonal projector onto
the orthogonal complement of ${\mathbf{A}}( {\bm{\Theta}}_0)$ in $\mathbb{C}^{M_TM_R}$, and $\bm{P}(\bm{\Theta}, \bm{\Phi})=\bm P_1$ has the same meaning with respect to $\mathbf{A}(\bm{\Theta},\bm{\Phi})$,  $\lambda_G$ is the detection threshold.
We underline in passing that this result can be easily obtained starting from the GLR
\begin{eqnarray}
	\label{eq:Test}
	\mathcal{T}=\frac{\max _{{\bm{\beta}},\sigma^2} p_1(\bm z \mid_{{ {\bm{\Theta}},{\bm{\Phi}},} {\bm{\beta}},\sigma^2} ;{\cal H}_1)}{\max _{ \bm{\alpha},\sigma^2} 	p_0(\bm z \mid_{{ \bm{\Theta}_0,} \bm{\alpha},\sigma^2 };{\cal H}_0)} 
\end{eqnarray}
with
\begin{eqnarray}
	\label{eq:p1}
	p_1(\bm z \mid_{  {\bm{\beta}},\sigma^2} ;{\cal H}_1) =\frac{1}{(\pi{\sigma^2})^{M_T M_R}}e^{-\frac{1}{{\sigma^2}} \|(\bm z-\mathbf{A({\bm{\Theta}},{\bm{\Phi}}){\bm{\beta}}}) \|_2^2 },
\end{eqnarray}
\begin{eqnarray}
	\label{eq:p0}
	p_0(\bm z \mid_{  \bm{\alpha},\sigma^2 };{\cal H}_0) =\frac{1}{(\pi{\sigma^2})^{M_T M_R}}e^{-\frac{1}{{\sigma^2}} \|(\bm z-\mathbf{A(\bm{\Theta}_0)\bm{\alpha}}) \|_2^2 }.
\end{eqnarray}
Since the Maximum-Likelihood estimates of $\bm \alpha$
and $\bm \beta$ are $ \mathbf{A^\dagger(\bm{\Theta}_0)}\bm z$ and 
$
\mathbf{A^\dagger({\bm{\Theta}},{\bm{\Phi}})}\bm z$, respectively, the final maximization steps of the numerator and the denominator of \eqref{eq:Test} with respect to $\sigma^2$ yields the test \eqref{eq:GLRT_un}.

The test \eqref{eq:GLRT_un}, which we adopt outright, complies with the prior constraints 1) and 2). Concerning 1), indeed, the test is {\em invariant} to transformations that rotate the observations in the range span of $\bm G=\bm P_0 \bm H$ and non-negatively {\em scale} $\bf z$ \cite{Scharf}. As we'll be shortly verifying, this results in a detection threshold and a false alarm probability which are independent of both $\mathbf{A(\bm{\Theta}_0)}$ and the noise floor $\sigma^2$. Concerning optimality, the test statistic in \eqref{eq:GLRT_un} turns out to be a {\em maximal invariant} statistic \cite{Lehmann}, whereby the test \eqref{eq:GLRT_un} is  Uniformly Most Powerful (UMP) one under the said invariance constraints.
\color{black}
\subsection{Performance Bounds Under Perfect Angle Estimation}
\label{sec:perfor_the}
In this part, we provide a statistical analysis of the proposed GLRT detector and derive expressions of  false alarm and detection probabilities.
Since $\mathbf{A({\bm{\Theta}},{\bm{\Phi}})}$ is a concatenation of $\mathbf{A(\bm{\Theta}_0)}$ with some $\bm E $, we have that ${\cal R}(\mathbf{A(\bm{\Theta}_0)})\subseteq {\cal R}(\mathbf{A({\bm{\Theta}},{\bm{\Phi}})})$, whereby 
\begin{eqnarray}
	\mathcal{R}(\bm{P}_1) \subseteq \mathcal{R}({\bm{P}}_0),
\end{eqnarray}
i.e.
\begin{eqnarray}
	\mathcal{R}({\bm{P}_0})=\mathcal{R}(\bm{P}_1) \oplus \bm{S}_{\perp},
\end{eqnarray}
where $\bm{S}_{\perp}$ denotes the orthogonal complement of $\mathcal{R}(\bm{P}_1)$ in $\mathcal{R}( {\bm{P}}_0)$. Denoting $\bm{P}^{  \bm{S}_{\perp} }$ as the orthogonal projector onto $ \bm{S}_{\perp}$, \color{black} and assuming that the echo signals from different paths are incoherent, we have $\text{dim}\left(\mathcal{R}\left(\boldsymbol{P}_1\right)\right)=M_T M_R-K_0 -2 K_1$, $\text{dim}\left(\mathcal{R}\left(\boldsymbol{P}_0\right)\right)=M_T M_R-K_0 $ and $\text{dim}\left(\mathcal{R}\left(\boldsymbol{P}^{\bm{S}_{\perp}}\right)\right)=2 K_1$. The test can be rewritten as 
\begin{eqnarray}
	\label{eq:test}
	\frac{\left\|\bm{P}_0 \bm{z}\right\|^2}{\left\|\bm{P}_1 \bm{z}\right\|^2}= \frac{\left\|\bm{P}_1 \bm{z}\right\|^2+\left\|\bm{P}^{\bm{S}_{\perp}}\bm{z}\right\|^2}  {\left\|\bm{P}_1 \bm{z}\right\|^2}
	=1 + \frac{ \left\|\bm{P}^{\bm{S}_{\perp}}\bm{z}\right\|^2}  {\left\|\bm{P}_1 \bm{z}\right\|^2} .
\end{eqnarray}
Defining $X =   \frac{\left\|\bm{P}^{\bm{S}_{\perp}}\bm{z}\right\|^2}  {\left\|\bm{P}_1 \bm{z}\right\|^2}$, under ${\cal H}_0$, we have 
$
{\left\|\bm{P}_0 {\mathbf{A}}(\bm{\Theta}_0)\bm{\alpha}\right\|^2 } = {\left\|\bm{P}_1 {\mathbf{A}}(\bm{\Theta}_0)\bm{\alpha}\right\|^2 }+ {\left\|\bm{P}^{\bm{S}_{\perp}}{\mathbf{A}}(\bm{\Theta}_0)\bm{\alpha}\right\|^2 } = 0
$, whereby
\begin{eqnarray}
	X =   \frac{\left\|\bm{P}^{\bm{S}_{\perp}}\bm{z}\right\|^2}  {\left\|\bm{P}_1 \bm{z}\right\|^2} =\frac{ \left\|  \bm{P}^{\bm{S}_{\perp}}\mathbf{w} \right\|^2}   { \left\|  \bm{P}_1\mathbf{w} \right\|^2}.
\end{eqnarray}  
Since $\mathbf{ w}\sim {\cal CN} (\mathbf{0}, \sigma^2 \bm I _{M_TM_R})$, under ${\cal H}_0$ the random variable $X$
is the ratio of two independent central Chi-square random variables, with $4K_1$ and $2(M_TM_R-K_0-2K_1)$ degrees of freedom, respectively, and hence has a Fisher-Snedecor distribution  with density
\begin{eqnarray}
	\label{eq:pro_h0}
	f_{X \mid {\cal H}_0}(x)=\frac{ 1}{B\left(2 K_1 ; m\right)} x^{2 K_1-1} (1+x)^{-(m+2K_1 )},
\end{eqnarray}
where $m =M_TM_R-K_0-2K_1 $ and ${B\left(a ; b\right)}$ denotes the beta function with parameters $a$ and $b$. 

\begin{figure}
		\centering
		\includegraphics[width=3in]{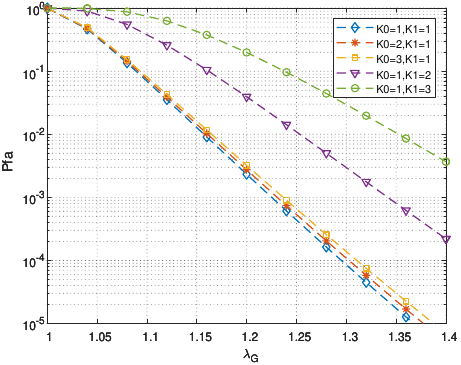}
		\caption{   $P_{\mathrm{fa}}$ versus $\lambda_G$ with $M_T M_R = 48$.}
		\label{fig:Pfa_difK_48}
	\end{figure}

In order to determine the density under ${\cal H}_1$, a model for ${\bm \beta}$ is to be chosen. A customary assumption is that ${\bm \beta} \sim C{\cal N}(0, \bm K_\beta)$, namely that it is a proper complex Gaussian vector with covariance matrix $\bm K_\beta$, which implies that the test statistic has again a Fisher-Snedecor distribution. Since
\begin{eqnarray}
	\mathbb{E}\left( { \left\|\bm{P}^{\bm{S}_{\perp}}\bm{z}\right\|^2}  |{\cal H}_1\right)  = \mathbb{E}\left( \left\|\bm{P}^{\bm{S}_{\perp}}{\mathbf{A}}(\bm{\Theta},\bm{\Phi}) \bm{\beta}  + \bm{P}^{\bm{S}_{\perp}}\mathbf{w} \right\|^2\right)\nonumber\\
	=\text{Trace}\left({\mathbf{A}}^H(\bm{\Theta},\bm{\Phi})\bm{P}^{\bm{S}_{\perp}}{\mathbf{A}}(\bm{\Theta},\bm{\Phi})\bm K_\beta \right) + \sigma^2 2 K_1.
\end{eqnarray}
We thus have that, under ${\cal H}_1$, the random variable $X$ has density
	\begin{eqnarray}
		\label{eq:H1_pro}
		f_{X \mid {\cal H}_1}(x)
		&=&\frac{1+\rho_1}{B\left(2 K_1 ; m\right)}\left(\frac{x }{1+ \rho_1}\right)    ^{2 K_1-1}\nonumber\\
		&&\times \left(1 + \frac{x}{1+\rho_1} \right)^{-(m+2K_1 )},
	\end{eqnarray}
	where 
	\begin{eqnarray}
		\rho_1 
		&=&  \frac{ \text{Trace}\left({\mathbf{A}}^H({\bm{\Theta}},{\bm{\Phi}})\bm{P}^{\bm{S}_{\perp}}{\mathbf{A}}({\bm{\Theta}},{\bm{\Phi}})\bm K_\beta\right)}{2K_1\sigma^2}.
	\end{eqnarray}
	Elementary calculations allow thus to determine the performance of the test in the form:
	\begin{eqnarray}
		\label{eq:Pfa}
		P_{\mathrm{fa}}&=& 1-	\int_{0 }^{\lambda_G -1} f_{X \mid {\cal H}_0}(x) d x \nonumber\\
		&=& 1- \frac{1}{B\left(2 K_1 ; m\right)} \sum_{i=0}^{m-1}\tbinom{m-1}{i}\nonumber\\
		&&\times	\frac{(-1)^i}{2 K_1+i}\left(1-\frac{1}{\lambda_G}\right)^{2 K_1+i},
	\end{eqnarray}
	
	\begin{eqnarray}
		\label{eq:pd_the}
		P_{\text{d}}&=& 1-	\int_{0 }^{\lambda_G -1} f_{X \mid {\cal H}_1}(x) d x \nonumber\\
		&=& 1- \frac{1}{B\left(2 K_1 ; m\right)} \sum_{i=0}^{m-1}\tbinom{m-1}{i}\nonumber\\
		&&\times\frac{(-1)^i}{2 K_1+i}\left(\frac{\lambda_G -1}{\lambda_G + \rho_1}\right)^{2 K_1+i}.
	\end{eqnarray}		
	\color{black} 	To analyze the impact of different $K_0$ and $K_1$  values on the theoretical performance, we conduct  experiments using a MIMO radar with $M_T = 6$ and $M_T = 8$.
	As far as the false alarm performance is concerned, we recall here that the test \eqref{eq:GLRT_un} achieves CFARness, whereby $P_{\mathrm{fa}}$ only depends on $K_0$ and $K_1$: sample plots of the behavior of $P_{\mathrm{fa}}$ versus the threshold for some values of $K_0$ and $K_1$ are reported in Fig.  \ref{fig:Pfa_difK_48}.
	\color{black}
	\begin{figure}
		\centering
		\includegraphics[width=3in]{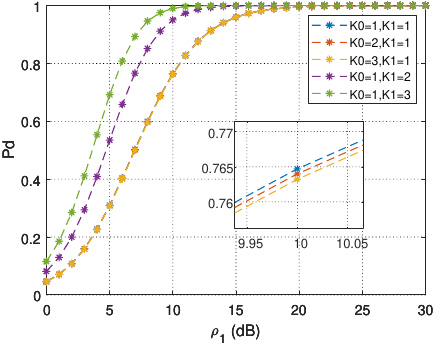}
		\caption{   $P_{\mathrm{d}}$ versus $\rho_1$ with $M_T M_R = 48$.	}
		\label{fig:Pd_difK_48}
	\end{figure} 
	\begin{figure}
		\centering
		\includegraphics[width=3in]{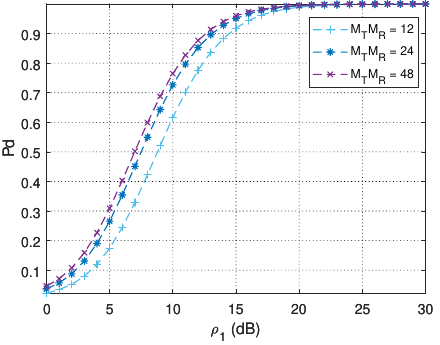}
		\caption{   $P_{\mathrm{d}}$ versus $\rho_1$ with $K_0 =1$, $K_1 = 1$.}
		\label{fig:Pd_difM}
	\end{figure} 
	
	Concerning the detection probability, we refer to the 
	interesting \color{black} special case that $\bm{\beta} \sim  {\cal CN} (0, \sigma_{\beta}^2\bm{I}_{2K_1})$, $\bm{\alpha} \sim  {\cal CN} (0, \sigma_{\alpha}^2\bm{I}_{K_0})$,  \color{black} which results into a  signal-to-noise ratio (SNR)
	\begin{eqnarray}
		\label{eq:rho_ideal}
		{\rho_1} &=& \frac{\sigma_{\beta}^2 \text{Trace}\left({\mathbf{A}}^H({\bm{\Theta}},{\bm{\Phi}})\bm{P}^{\bm{S}_{\perp}}{\mathbf{A}}({\bm{\Theta}},{\bm{\Phi}})\right)}{2K_1\sigma ^2}\nonumber\\
		&=& \frac{\sigma_{\beta}^2 2 K_1}{2K_1\sigma ^2} = \frac{\sigma_{\beta}^2}{\sigma ^2}.
	\end{eqnarray}
	Fig.  \ref{fig:Pd_difK_48} and \ref{fig:Pd_difM} highlight the detector behavior for different values of $(K_0,K_1)$ and the impact of the number of system degrees of freedom $M_TM_R$. Not surprisingly, we observe that larger values of $K_1$ for fixed $K_0$ result in better detection performance. This is obviously because the two subspaces defined by the projection matrices $\bm P_0$ and $\bm P_1$ become more and more distinguishable as $K_1$ increases: the inevitable consequence is that the "worst case" is the situation where $K_0$ is large (in the plot, $K_0=3$) and $K_1$ small (in the plot, $K_1=1$). As to the influence of $M_TM_R$, the detection performance -as intuitively evident - improves with increasing number of degrees of freedom, even though the gain rapidly decreases once $M_TM_R$ is made sufficiently large as compared to the values of $K_0$ and $K_1$.

	{\section{Angle estimation for multipaths}}
	As anticipated, the test \eqref{eq:GLRT_un} is not implementable, in that the two matrices $\mathbf A(\bm \Theta_0)$ and $\mathbf A(\bm \Theta, \bm \Phi)$ are not known even in their order. In principle, such a prior uncertainty could be addressed within the GLRT framework. Noticing that directly solving the problems $\min_{K_0,\bm A(\bm \Theta_0) \in \mathbb{C}^{M_TM_R\times K_0}}
	\parallel \bm P (\bm \Theta_0)\bm z \parallel^2$ and $\min_{K_0,K_1,\bm A(\bm \Theta, \bm \Phi) \in 
		\mathbb{C}^{M_TM_R\times (K_0+2K_1)}}\parallel \bm P (\bm \Theta, \bm \Phi)\bm z \parallel^2$ leads to an overestimation of the model order, we consider introducing the sparsity of the reflection paths to estimate $\mathbf A(\bm \Theta_0)$ and $\mathbf A(\bm \Theta, \bm \Phi)$, i.e., $K_0$ and $K_1$ are usually much smaller than $M_T M_R $ \cite{yang_sparsity}.  Specifically, we try to resolve the following problem assuming ${\cal H}_0$ hypothesis is true:
			\begin{eqnarray}
			\label{eq:H0_problem}
			\begin{aligned}
				&(\hat{K}_0,\widehat{\bm \Theta}_0,\hat{ \bm \alpha}) = \mathop {\arg \min }\limits_{{\bm \Theta}_0\in \mathbb{R}^{K_0\times1},{\bm \alpha} \in \mathbb{C}^{K_0\times1},K_0}  K_0\\
				&\text { s.t. }\left\| {\bm z- {\mathbf{A}}({\bm{\Theta}}_0) \bm \alpha } \right\|_2^2 \leq\epsilon^2,			\end{aligned}
		\end{eqnarray}
		and obtain a suitable estimation of the direct paths $\widehat{\bm \Theta}_0$. It is worth noting that for any ${\bm \Theta}_0$, $ \bm \alpha $ should be $ A^\dagger(\bm \Theta_0) \bm z $ to minimize $\left\| {\bm z- {\mathbf{A}}({\bm{\Theta}}_0) \bm \alpha } \right\|_2^2 $, so the constraints degrades to $\parallel \bm P (\bm \Theta_0)\bm z \parallel_2^2 \leq\epsilon$.

		Similarly, assuming  ${\cal H}_1$  hypothesis is true, we resort to:
		\begin{eqnarray}
			\label{eq:H1_problem}
			\begin{aligned}
				&(\hat{K}_0,\hat{K}_1,\widehat{\bm \Theta},\widehat{\bm \Phi},\hat{ \bm \beta}) = \mathop {\arg \min }\limits_{\substack{{K}_0,{K}_1,\\ {\bm \Theta}\in \mathbb{R}^{(K_0+2K_1)\times1},\\{\bm \Phi}\in \mathbb{R}^{(K_0+2K_1)\times1},\\{\bm \beta} \in \mathbb{C}^{(K_0+2K_1)\times1}}}  {K_0+\delta K_1}\\
				&\text { s.t. }\left\| {\bm z- {\mathbf{A}}({\bm{\Theta}},\bm \Phi) \bm \beta } \right\|_2^2 \leq\epsilon^2,
			\end{aligned}
		\end{eqnarray}
		where $\delta$ is the parameter characterizing the weights between $K_0$ and $K_1$. 
	The test family is then applied for the detection of ${\cal H}_1$ from ${\cal H}_0$
	\begin{equation}
		\frac{\parallel (\bm I_{M_TM_R}-\mathbf A( \widehat{\bm \Theta}_0)\mathbf A^\dagger ( \widehat{\bm \Theta}_0)) \bm z\parallel^2}{\parallel (\bm I_{M_TM_R}-\mathbf A( \widehat{\bm \Theta},\widehat{\bm \Phi})\mathbf A^\dagger(\widehat{\bm \Theta},\widehat{\bm \Phi}) )\bm z\parallel^2} \underset{{\cal H}_0}{\stackrel{{\cal H}_1}{\gtrless}} \lambda_G.
	\end{equation}
	 {The next two subsections are thus devoted to illustrating how the needed estimators may be designed to solve \eqref{eq:H0_problem} and \eqref{eq:H1_problem} through bounded-complexity procedures. }
	\color{black}
	\medskip
	
	\subsection{Estimators for $\bm \Theta_0$ under ${\cal H}_0$ hypothesis }
 We propose an   iterative scheme to solve the problem given by \eqref{eq:H0_problem}. Define the residual in the $t$-th as ${ \bm r }^{(t)}$. At the beginning of the iterations, the residual is  initialized as  $\mathbf{r}^{(0)} = \bm z$ and  the angle set of the paths is initialized to be empty, i.e.,   $\widehat{\bm{\Theta}}^{(0)}_0= \emptyset  $ and $\hat K_0^{(0)}=0$. In the $t$-th iteration, we insert a path into the set and $\hat K_0^{(t)} $ is updated as $\hat K_0^{(t)}=\hat K_0^{(t-1)}+1$. The angle of the path is firstly selected from a set of grid points, then an iterative refinement in continuous domain is applied. 
 
 To minimize the $\ell_2$-norm of the residual as much as possible, we select the angle that has the largest correlation with  the residual $\mathbf{r}^{(t-1)}$ from a $G$ uniformly distributed grids $\{\tilde{\theta}_1,\tilde{\theta}_2,\dots,\tilde{\theta}_G\} $, i.e.,
   	 \begin{eqnarray}
 	\label{eq:addAtom_H0}
 	\hat{\theta}^{(t)} = \mathop{\arg \max}\limits_{\theta^{(t)} \in \{\tilde{\theta}_1,\tilde{\theta}_2, \cdots,\tilde{\theta}_ G \} } |(\mathbf{r}^{(t-1)})^H\bar{\mathbf{a}}(\theta^{(t)})|.
 \end{eqnarray}
The angle set is roughly estimated as
   \begin{eqnarray}
 	\label{eq:add_h0}
 	\widehat{\bm{\Theta}}^{(t,0)}_0 = [(\widehat{\bm{\Theta}}^{(t-1)}_0)^T, \hat{\theta}^{(t)}]^T.
 \end{eqnarray}
Here, the grid-search initialization helps to avoid the local optimum of directly solving \eqref{eq:H0_problem} in continuous domain. To improve the accuracy of the estimation, a refinement step is then implemented by solving
   	\begin{eqnarray}
 	\label{eq:F_0}
\widehat{\bm{\Theta}}_0^{(t)}&= &\mathop {\arg \min }\limits_{{\bm{\Theta}}_0^{(t)}\in \mathbb{R}^{\hat K_0^{(t)}\times1} }{F({\bm{\Theta}}_0^{(t)}) }\\ \nonumber
&=&\mathop {\arg \min }\limits_{{\bm{\Theta}}_0^{(t)}\in \mathbb{R}^{\hat K_0^{(t)}\times1} } \parallel \bm z-\mathbf A( {\bm \Theta}_0^{(t)})\mathbf A^\dagger ( {\bm \Theta}_0^{(t)})\bm z \parallel^2_2.
 \end{eqnarray}
We can use Gauss–Newton (GN)   method to find the optimal solution. Starting with $\widehat{\bm{\Theta}}^{(t,0)}_0$, the GN method updates  
	\begin{eqnarray}
		\label{eq:updata_theta0}
		\widehat{\bm{\Theta}}^{(t,i+1)}_0 = 	\widehat{\bm{\Theta}}_0^{(t,i)}+\mathbf{h}^{(t,i)}_0,
	\end{eqnarray}
	where 
	\begin{eqnarray}
		\label{eq:h_gn_0}
		\mathbf{h}_{0}^{(t,i)} = -(\mathbf{H}_0^{(t,i)})^{-1}\mathbf{g}_0^{(t,i)},
	\end{eqnarray}
	with $\mathbf{g}_0^{(t,i)}$ 	and  $\mathbf{H}_0^{(t,i)}$ denoting gradient and Hessian  of function $F({\bm{\Theta}}_0^{(t,i)}) $, respectively. Define  $\mathbf{A}^{(t,i)}_0 = \mathbf{A} ({\bm{\Theta}}_0^{(t,i)})$, $\bm{P}^{(t,i)}_0 =\bm{P}({\bm{\Theta}}_0^{(t,i)})$.  Following the derivations  in Appendix A,  the expressions of $\mathbf{H}_0^{(t,i)}$ and $\mathbf{g}_0^{(t,i)}$ are  given by 
	\begin{eqnarray}
		\label{eq:g0_h0}
		\mathbf{g}_0^{(t,i)} =-2{\rm{Re}}\left\{{\rm{\rm{diag}}}\{(\mathbf{A}^{(t,i)}_0)^\dagger \bm z\bm z^H  \bm{P}^{(t,i)}_0 \mathbf{D}_0^{(t,i)}\}\right\},
	\end{eqnarray}
	\begin{eqnarray}
		\label{eq:H0_h0}
		\begin{aligned}
			\mathbf{H}_{0}^{(t,i)}& =2{\rm{Re}}\left\{ (\mathbf{D}_0^{(t,i)})^H \bm{P}^{(t,i)}_0 \mathbf{D}_0^{(t,i)}\right.\\
			&\left.\odot \left( (\mathbf{A}^{(t,i)}_0)^\dagger \bm z \bm z^H((\mathbf{A}^{(t,i)}_0)^\dagger )^H\right)^T\right\}\\
			&+2{\rm{Re}}\left\{   (\mathbf{D}_0^{(t,i)})^H \bm{P}_0^{(t,i)}\bm z\bm z^H  \bm{P}_0^{(t,i)} (\mathbf{D}_0^{(t,i)})^T \right.\\
			&\left.\odot\left((\mathbf{A}^{(t,i)}_0)^\dagger((\mathbf{A}^{(t,i)}_0)^\dagger)^H\right) \right\}
		\end{aligned}
	\end{eqnarray}
	where 	$\mathbf{D}_{0}^{(t,i)} = \left[\frac{\partial {\mathbf{a}}(\hat{\theta }^{(i)}_1) }{\partial  	\hat{\theta }^{(i)}_1},\frac{\partial {\mathbf{a}}(\hat{\theta }^{(i)}_2) }{\partial  	\hat{\theta }^{(i)}_2},\dots,\frac{\partial {\mathbf{a}}(\hat{\theta }^{(i)}_t) }{\partial  	\hat{\theta }^{(i)}_{t}}\right]^T$ for $i=1,2,\dots,I$.
	
	\color{black}

	The above computations are carried out iteratively until  
	 a maximum iteration number $I$ is reached.  We take the solution $\widehat{\bm{\Theta}}_0^{(t,I)}$ as the refined estimated angle  $\widehat{\bm{\Theta}}_0^{(t)}$ in the $t$-th  iteration, and update the amplitude estimation  and residual as follows:
		 \begin{eqnarray}
			 	\hat{\bm \alpha}^{(t)}&=& \mathbf{A}^\dagger(\widehat{\bm \Theta}_0^{(t)})\bm z,\label{eq:up_alpha}\\
			 	\mathbf{r}^{(t)} & = &\bm z - \mathbf{A}(\widehat{\bm \Theta}_0^{(t)})\hat{\bm \alpha}^{(t)}.\label{eq:up_residual}
			 \end{eqnarray}
			 
 In principle, the iterative process stops when  $\|\mathbf{r}^{(t)}\|_2 \leq\epsilon$ i.e the condition of  \eqref{eq:H0_problem} is satisfied. Additionally, considering the potential issue of overestimating $K_0$, {we also set a maximum number of iterations 
$T$. The iterative process will be terminated if $t\geq T$ or $\|\mathbf{r}^{(t-1)}\|_2- \|\mathbf{r}^{(t)}\|_2 \leq \epsilon_1$.}
	The detailed procedure is given  in Algorithm \ref{al:NLS-GN-direct} and we name the proposed method as Compressed Sensing method in Continuous Domain under hypothesis  ${\cal H}_0$ (CSCD-H0) algorithm.
	\begin{algorithm}[h]
		\caption{{ CSCD-H0 algorithm}} 
		\label{al:NLS-GN-direct}
		\KwIn{${{\bm z}}$, $\{\tilde{\theta}_1,\tilde{\theta}_2,\dots,\tilde{\theta}_G\}$
			and $T$, $I$, $\epsilon $, $\epsilon_1$;} 
		\KwOut{$\hat{K}_0$, $\hat{\bm{\alpha}}\in \mathbb{C}^{\hat{K}_0\times1}$, $\widehat{\bm{\Theta}}_0\in \mathbb{R}^{\hat{K}_0\times1}$;}
		{\bf Initialization:} $\widehat{\bm{\Theta}}^{(0)}_0= \emptyset  $ and   $\hat{K}_0^{(0)} =0,\mathbf{r}^{(0)}= \bm z, t = 0;$\\
		\While{ $\|\mathbf{r}^{(t)}\|_2 > \epsilon$  }
		{ 
			$t\leftarrow  t+1$;\\
			Obtain the inserted angle $\hat{\theta}^{(t)}$ via   \eqref{eq:addAtom_H0};\\
			$	\widehat{\bm{\Theta}}^{(t,0)}_0 = [(\widehat{\bm{\Theta}}^{(t-1)}_0)^T, \hat{\theta}^{(t)}]^T$, 	$\hat{K}_0^{(t)} = \hat{K}_0^{(t-1)}+1$, ;\\
			\For{ $i = 0$ to  $I$ }
			{
				Calculate  $\mathbf{g}_0^{(t,i)}$ and  $\mathbf{H}_0^{(t,i)}$ according to   \eqref{eq:g0_h0} and    	\eqref{eq:H0_h0},  respectively; \\
				Calculate $\mathbf{h}_{0}^{(t,i)}$ by \eqref{eq:h_gn_0} and update $\widehat{\bm{\Theta}}_0^{(t,i+1)}$ by \eqref{eq:updata_theta0};\\
			}
			 Update $\widehat{\bm{\Theta}}_0^{(t)} \leftarrow \widehat  {\bm{\Theta}}_0^{(t,I)}$, \\
			Update $\hat{\bm \alpha}^{(t)}$  and residue $\mathbf{r}^{(t)}  $ by \eqref{eq:up_alpha} and \eqref{eq:up_residual}, respectively;\\
			\If{  $t \geq T$ $\rm or$ $\|\mathbf{r}^{(t-1)}\|_2- \|\mathbf{r}^{(t)}\|_2 \leq \epsilon_1$ }{Break;}		 		 	
		}
		Return: $\widehat{\bm{\Theta}}_0= \widehat{\bm{\Theta}}^{(t)}_0$,  $\hat{\bm{\alpha}} =  \hat{\bm \alpha}^{(t)}$, $\hat{K}_0 = \hat{K}_0^{(t)}$ .
	\end{algorithm}
	
It is worth noting that, without the refinement step, the algorithm degrades to the Orthogonal Matching
Pursuit (OMP) algorithm \cite{OMPforDOA}, which is  a classic method in CS. As OMP doesn't involve grid refinement in the continuous domain, the computational complexity is lower. However, the algorithm   suffers from the off-grid problem and is likely  to overestimate the value of  $K_0$.
	
	\subsection{Estimators for $ (\bm\Theta, \bm \Phi)$ under ${\cal H}_1$ hypothesis }
	
	Assuming ${\cal H}_1$ hypothesis is true, we need to estimate the angle of both direct   and first-order paths. 	 The parameters are initialized as ${\bf r }^{(0)} = {\bf z }, \widehat{\bm \Theta}_1^{(0)} = \emptyset,\widehat{\bm \Phi}_1^{(0)} = \emptyset,\widehat{\bm \Theta}_0^{(0)} = \emptyset$, $\hat K _1^{(0)}=0$, $\hat K _0^{(0)}=0$. In the $t$-th iteration, we adopt two individual processes to estimate the angles, respectively assuming the direct  and first-order paths should be taken. Then, we check the residual of both cases and decide the type of paths to keep  for the $t$-th iteration.
 
	 In  case 1 where a direct path is taken,  the angle is selected according to \eqref{eq:addAtom_H0} and the rough estimation of angle set $\bar{\bm{\Theta}}^{(t,0)} =[\bar{\bm{\Theta}}^{(t,0)}_1 ;\bar{\bm{\Phi}}^{(t,0)}_1 ;\bar{\bm{\Theta}}^{(t,0)}_0 ]$ is obtained with $	{\bar{\bm{\Theta}}} ^{(t,0)}_1 = \widehat{\bm{\Theta}}^{(t-1)}_1, 
	 {\bar{\bm{\Phi}}}^{(t,0 )}_1  = \widehat{\bm{\Phi}}^{(t-1)}_1$ and $
	 {\bar{\bm{\Theta}}}^{(t,0 )}_0  = [(\widehat{\bm{\Theta}}^{(t-1)}_0)^T,\hat{\theta}^{(t)}]^T$.
	 
	 In  case 2 where  a pair of first-order paths are added, the  angle pair {$(\hat{ \vartheta}^{(t)},\hat{ \varphi}^{(t)} )$ } are taken  from two $G$ uniformly distributed grids   $\bm{\Xi}_t = \{\tilde{\vartheta}_1,\tilde{\vartheta}_2,\dots,\tilde{\vartheta}_G\}$ and $ \bm{\Xi}_r= \{\tilde{\varphi}_1,\tilde{\varphi}_2,\dots,\tilde{\varphi}_G\}$ by
	 \begin{equation}
	 	\label{eq:addAtom_H1}
	 	\begin{split}
	 		(\hat{ \vartheta}^{(t)},\hat{ \varphi}^{(t)} ) 
	 		=  &	 \mathop{\arg 	\max} \limits_{\substack{{\vartheta}^{(t)}\in  \bm \Xi _t \\ {\varphi}^{(t)}\in  \bm \Xi _r\\ {\vartheta}^{(t)}<{\varphi}^{(t)}}}
	 		\left(|(\mathbf{r}^{(t-1)})^H(\mathbf{a}_T({\vartheta}^{(t)})\circ \mathbf{a}_R({\varphi}^{(t)}))|\right.  \\  
	 		&\left.+|(\mathbf{r}^{(t-1)})^H(\mathbf{a}_T({\varphi}^{(t)})\circ \mathbf{a}_R({\vartheta}^{(t)}))|\right).
	 	\end{split}
	 \end{equation} 
	 The angles $\bar{\bar{\bm{\Theta}}}^{(t,0)} =[\bar{\bar{\bm{\Theta}}}^{(t,0)}_1 ;\bar{\bar{\bm{\Phi}}}_1^{(t,0)} ;\bar{\bar{\bm{\Theta}}}_0^{(t,0)} ]$   with  $	\bar{\bar{\bm{\Theta}}} ^{(t,0)}_1 = [(\widehat{\bm{\Theta}}^{(t-1)}_1)^T,\hat{\vartheta}^{(t)}]^T, 
	 \bar{\bar{\bm{\Phi}}}^{(t,0 )}_1 = [(\widehat{\bm{\Phi}}^{(t-1)}_1)^T,\hat{\varphi}^{(t)}]^T$ and $
	 \bar{\bar{\bm{\Theta}}}^{(t,0 )}_0  = \widehat{\bm{\Theta}}^{(t-1)}_0$ are roughly estimated.
	  

Since the  rough estimation of angle set  above  is obtained based on grid search,  we further need to employ optimization methods to achieve angle estimation in the continuous domain.
Due to the mixture of direct and first-order paths under the ${\cal H}_1$ hypothesis,  the GN method may lead to unstable estimation due to the {rank-deficiency} in Hessian. Therefore, we resort to Levenberg-Marquardt (LM) method \cite{Gavin2013TheLM} for updating angle estimates.
		 
{ For abbreviation, here we show how this LM iteration works based on the setting of case 1. For case 2, the process follows the same flow just  with  $ \bar{\bm{\Theta}}^{(t,0)}$   replaced by $\bar{\bar{\bm{\Theta}}}^{(t,0)}$. 
 The angles set is updated as  $\bar{\bm{\Theta}}^{(t,i+1)}=\bar{\bm{\Theta}}^{(t,i)}
		+ \mathbf{h}^{(t,i)}$, }
	where
		\begin{eqnarray}
			\label{eq:update_lm}
			\mathbf{h}^{(t,i)} = -(\mathbf{H}^{(t,i)}+\mu^{(t,i)}\bm{I}_{\hat{K}^{(t)}})^{-1}\mathbf{g}^{(t,i)},
		\end{eqnarray}
	 with $\mathbf{H}^{(t,i)}$  and $\mathbf{g}^{(t,i)}$ denoting the Hessian and gradient  of  $\bar F(\bar{\bm{\Theta}}^{(t,i)}) =  \parallel \bm z-\mathbf A( \bar{\bm \Theta}^{(t,i)}, \bar{\bm \Phi}^{(t,i)})\mathbf A^\dagger ( \bar{\bm \Theta}^{(t,i)}, \bar{\bm \Phi}^{(t,i)})\bm z \parallel^2_2 $, respectively. $\hat{K}^{(t)}$ denotes the size of $\bar{\bm{\Theta}}^{(t,i)}$,  $\mu^{(t,i)}$ is a damping parameter. 
		We want to emphasize that the calculations of $\mathbf{g}^{(t,i)} $ and $\mathbf{H}^{(t,i)} $ are different from those under the ${\cal H}_0$ hypothesis.   Let us partition  $\mathbf{g}^{(t,i)} $ into:
		\begin{eqnarray}
			\label{eq:g_group}
			\mathbf{g}^{(t,i)} = \left[\mathbf{g}_{\rm T}^{(t,i)} ;\mathbf{g}_{\rm R}^{(t,i)};{\mathbf{g}'_0}^{(t,i)} \right],
		\end{eqnarray} 
		where $\mathbf{g}_{\rm T}^{(t,i)}$ and $ \mathbf{g}_{\rm R}^{(t,i)}$ denote  the gradient of $\bar F$  with respect to  DOD's and DOA's of first-order paths, respectively,  ${\mathbf{g}'_0}^{(t,i)}$ denotes  the gradient of $\bar F$  with respect to the  DOA's of direct paths. The specific forms of elements $\mathbf{g}$'s  are given  by \eqref{eq:g_DOD} -\eqref{eq:g_dir} in Appendix \ref{gradient_hessian_deriv}. Similarly,   $\mathbf{H}^{(t,i)} $ is given by
		\begin{eqnarray}
			\label{eq:H_group}
			\mathbf{H}^{(t,i)} = \left[	\begin{array}{cccccc}
				\mathbf{H}_{\rm TT}^{(t,i)} &\mathbf{H}_{\rm TR}^{(t,i)}&\mathbf{H}_{\rm T0}^{(t,i)} \\ 
				\mathbf{H}_{\rm RT}^{(t,i)} &\mathbf{H}_{\rm RR}^{(t,i)}&\mathbf{H}_{\rm R0}^{(t,i)} \\
				\mathbf{H}_{\rm 0T}^{(t,i)} &\mathbf{H}_{\rm 0R}^{(t,i)}&\mathbf{H}_{00}^{(t,i)} \\
			\end{array}\right],
		\end{eqnarray}
	where the forms of elements  $\mathbf{H}$'s are given  by   \eqref{eq:HTT}-\eqref{eq:H00}  in Appendix \ref{gradient_hessian_deriv}.

	 In our algorithm, the paths with non-equivalent DOD and DOA are added in pairs. This aligns with the real-world scenario where the   first-order paths always appear in a paired, group-sparse manner. When calculating derivatives in $\mathbf{g}^{(t,i)} $ and $\mathbf{H}^{(t,i)} $, the pairwise constraint of the first-order paths must be considered.
	 For instance,  when  calculating the derivative of $\bar F$ with respect to  $\hat{ \vartheta}^{(t)}$, 
		the derivative of both  $\mathbf{a}_T(\hat{ \vartheta}^{(t)})\circ  \mathbf{a}_R(\hat{ \varphi}^{(t)})$ and $\mathbf{a}_T(\hat{ \varphi}^{(t)})\circ \mathbf{a}_R(\hat{ \vartheta}^{(t)})$ should be calculated.  These enforcements allow us  to leverage the group-sparsity of the first-order paths to enhance the estimation accuracy.

   The damping parameter $\mu^{(t,i)}$ in \eqref{eq:update_lm} is selected by a line search algorithm that is controlled by the gain ratio
		\begin{eqnarray}
			\label{eq:varrho}
			\varrho^{(t,i)} 	&=& \frac{ \bar F(\bar{\bm{\Theta}}^{(t,i)} )- \bar F(\bar{\bm{\Theta}}^{(t,i)}+ \mathbf{h}^{(t,i)})}{\frac{1}{2}\mathbf{h}^{(t,i)} ( \mu^{(t,i)} \mathbf{h}^{(t,i)} -\mathbf{g}^{(t,i)})}.
		\end{eqnarray}
		Steps 9-14 in Algorithm 2 describes how this parameter is obtained.

		After the iterative optimization of the angles, we obtain the angle set  $\bar{\bm{\Theta}}^{(t,T)} = [\bar{\bm{\Theta}}^{(t ,I)}_1 ;\bar{\bm{\Phi}}^{(t ,I)}_1;\bar{\bm{\Theta}}^{(t,I)}_0 ]$ and  the  residual  ${\bm r}_1^{(t)}$ for case 1, and   $\bar{\bar{\bm{\Theta}}}^{(t,T)} = [\bar{\bar{\bm{\Theta}}}^{(t ,I)}_1 ;\bar{\bar{\bm{\Phi}}}^{(t ,I)}_1;\bar{\bar{\bm{\Theta}}}^{(t,I)}_0 ]$ and  ${\bm r}_2^{(t)}$ for case 2.
We use the residual to select the type of output angles in the $t$-th iteration, i.e. when { $ \|{\bm r}_2^{(t)}\|_2 - \|{\bm r}_1^{(t)}\|_2<\delta_r$}, where $\delta_r$ is the parameter affecting the selection direct path or first-order path pair, 
 a pair of first-order paths is added, the angle estimation is updated as
\begin{eqnarray}
	\label{eq:add_fir}
	(\widehat{\bm{\Theta}}^{(t)}_1 ,\widehat{\bm{\Phi}}^{(t)}_1,\widehat{\bm{\Theta}}^{(t)}_0 )& =& (\bar{\bm{\Theta}}^{(t ,I)}_1,\bar{\bm{\Phi}}^{(t ,I)}_1,\bar{\bm{\Theta}}^{(t,I)}_0 ),
\end{eqnarray}
Otherwise, we have
\begin{eqnarray}
	\label{eq:add_dir}
	(\widehat{\bm{\Theta}}^{(t)}_1 ,\widehat{\bm{\Phi}}^{(t)}_1,\widehat{\bm{\Theta}}^{(t)}_0 )& =& (\bar{\bar{\bm{\Theta}}}^{(t ,I)}_1 ,\bar{\bar{\bm{\Phi}}}^{(t ,I)}_1,\bar{\bar{\bm{\Theta}}}^{(t,I)}_0 ).
\end{eqnarray}

The proposed algorithm is named as the   Compressed Sensing method in Continuous Domain under hypothesis  ${\cal H}_1$ (CSCD-H1) and  has been summarized  in Algorithm \ref{al:NLS-LM-group}. Similarly, when we remove the angle refinement in continuous domain and  just select paths  based on the maximum correlation, the method reduces to  a kind of group OMP (GOMP) estimation technique.
\begin{algorithm}
	\caption{{CSCD-H1 algorithm}} 
	\label{al:NLS-LM-group}
	\KwIn{
		$\bm z$, $\bm{\Xi}_t$, $\bm{\Xi}_r$
		and $T$, $I$, $J$,$\epsilon$,$\epsilon_2$, $\delta_r$ ;}
	\KwOut{   $\hat{K}_1$, $\hat{K}_0$, $\widehat{{\bm{\Theta}}}$, $\widehat{{\bm{\Phi}}}$, $\hat{\bm{\beta}}$;}
	{\bf Initialization:}  $\widehat{\bm{\Theta}}^{(0)}_1= \emptyset $, $\widehat{\bm{\Phi}}^{(0)}_1= \emptyset $, $\widehat{\bm{\Theta}}^{(0)}_0= \emptyset $,   $\mathbf{r}^{(0)}= \bm z$, $\hat{K}_0^{(0)} = 0$,     $\hat{K}_1^{(0)} = 0$, $t=0$;\\
	\While{ $\|\mathbf{r}^{(t)}\|_2 > \epsilon$ }
	{
		$t\leftarrow  t+1$;\\
			Obtain the  angle of direct path $\hat{\theta}^{(t)}$ via   \eqref{eq:addAtom_H0};\\
			Obtain $\bar{\bm{\Theta}}^{(t,0)} =[\bar{\bm{\Theta}}^{(t,0)}_1 ;\bar{\bm{\Phi}}^{(t,0)}_1 ;\bar{\bm{\Theta}}^{(t,0)}_0 ]$ where $	{\bar{\bm{\Theta}}} ^{(t,0)}_1 = \widehat{\bm{\Theta}}^{(t-1)}_1, 
			{\bar{\bm{\Phi}}}^{(t,0 )}_1  = \widehat{\bm{\Phi}}^{(t-1)}_1$ and $
			{\bar{\bm{\Theta}}}^{(t,0 )}_0  = [(\widehat{\bm{\Theta}}^{(t-1)}_0)^T,\hat{\theta}^{(t)}]^T$;\\
			\For{ $i = 0$ to  $I$ }
		 {
		 	Calculate  $\mathbf{g} ^{(t,i)}$ and  $\mathbf{H} ^{(t,i)}$ using   \eqref{eq:g_group} and    	\eqref{eq:H_group},  respectively; \\
		 	Calculate $\mathbf{h}^{(t,i)}$ and $\varrho^{(t,i)}$  by \eqref{eq:update_lm}  and \eqref{eq:varrho}, respectively;\\
		     $j\leftarrow0$;\\
			\While{$\varrho^{(t,i)} \leq 0$ $\rm and$ $j<J$}
			{Update $j \leftarrow j+1 $, $\mu^{(t,i)}\leftarrow2^j\mu^{(t,i)}$;\\
					Calculate $\mathbf{h}^{(t,i)}$ and $\varrho^{(t,i)}$  by \eqref{eq:update_lm}  and \eqref{eq:varrho}, respectively;}
			$\mu^{(t,i+1)} =\mu^{(t,i)} \max \{\frac{1}{3}, 1-    (2\varrho^{(t,i)} -1)^3\} $;\\ 
				$\bar{\bm{\Theta}}^{(t,i+1)}=  \bar{\bm{\Theta}}^{(t,i)} 	+ \mathbf{h}^{(t,i)}$;\\
		 	}
		 
		 ${\bm r}_1^{(t)} = \bm z - \mathbf{A}(\bar{\bm \Theta}^{(t,I)},\bar{\bm \Phi}^{(t,I)})\mathbf{A}^\dagger(\bar{\bm \Theta}^{(t,I)},\bar{\bm \Phi}^{(t,I)})\bm z $;\\
			{Obtain the  inserted angle pair} $(\hat{\vartheta}^{(t)},\hat{\varphi}^{(t)})$ via \eqref{eq:addAtom_H1};\\
			Obtain $\bar{\bar{\bm{\Theta}}}^{(t,0)} =[\bar{\bar{\bm{\Theta}}}^{(t,0)}_1 ;\bar{\bar{\bm{\Phi}}}_1^{(t,0)} ;\bar{\bar{\bm{\Theta}}}_0^{(t,0)} ]$ where $	\bar{\bar{\bm{\Theta}}} ^{(t,0)}_1 = [(\widehat{\bm{\Theta}}^{(t-1)}_1)^T,\hat{\vartheta}^{(t)}]^T, 
			\bar{\bar{\bm{\Phi}}}^{(t,0 )}_1 = [(\widehat{\bm{\Phi}}^{(t-1)}_1)^T,\hat{\varphi}^{(t)}]^T$ and $
			\bar{\bar{\bm{\Theta}}}^{(t,0 )}_0  = \widehat{\bm{\Theta}}^{(t-1)}_0$;\\
			
	 Optimize the	$\bar{\bar{\bm{\Theta}}}^{(t,0)}$ based on  LM method  given by step 6 to step 16 { with $ {\bar{\bm{\Theta}}}^{(t,0)}$ replaced by $\bar{\bar{\bm{\Theta}}}^{(t,0)}$} ;\\
		 	 ${\bm r}_2^{(t)} = \bm z - \mathbf{A}(\bar{\bar{\bm{\Theta}}}^{(t,I)},\bar{\bar{\bm{\Phi}}}^{(t,I)})\mathbf{A}^\dagger(\bar{\bar{\bm{\Phi}}}^{(t,I)},\bar{\bar{\bm{\Theta}}}^{(t,I)})\bm z $;\\
		\eIf{ $ \|{\bm r}_2^{(t)}\|_2 - \|{\bm r}_1^{(t)}\|_2<\delta_r$}
		{Update $\widehat{\bm{\Theta}}^{(t)}_1 ,\widehat{\bm{\Phi}}^{(t)}_1,\widehat{\bm{\Theta}}^{(t)}_0 $ by  \eqref{eq:add_fir};\\
			Update ${\bm r}^{(t)} = {\bm r}_2^{(t)}$;\\	  	
		Update $\hat{K}_0^{(t)} =  \hat{K}_0^{(t-1)}$, $\hat{K}_1^{(t)}  =  \hat{K}_1^{(t-1)}  +1$;}
		{  Update $\widehat{\bm{\Theta}}^{(t)}_1 ,\widehat{\bm{\Phi}}^{(t)}_1,\widehat{\bm{\Theta}}^{(t)}_0 $ by  \eqref{eq:add_dir};\\
			Update ${\bm r}^{(t)} = {\bm r}_1^{(t)}$;\\
			Update	  $\hat{K}_0^{(t)} =  \hat{K}_0^{(t-1)}  +1$, $\hat{K}_1^{(t)}  =  \hat{K}_1^{(t-1)}$.} 
		\If{  $t \geq T$ $\rm or$ $\|\mathbf{r}^{(t-1)}\|_2- \|\mathbf{r}^{(t)}\|_2 \leq \epsilon_2$ }{Break;}
	}
	Return: $ \widehat{\bm{\Theta}}  = [\widehat{\bm{\Theta}}^{(t)}_1 ,\widehat{\bm{\Phi}}^{(t)}_1,\widehat{\bm{\Theta}}^{(t)}_0]$, $\widehat{{\bm{\Phi}}}  = [\widehat{\bm{\Phi}}^{(t)}_1 ,\widehat{\bm{\Theta}}^{(t)}_1,\widehat{\bm{\Theta}}^{(t)}_0]$,  $\hat{\bm{\beta}} =  \mathbf{A}^\dagger (\widehat{\bm{\Theta}},\widehat{\bm{\Phi}}){{\bm z}}$, $\hat{K}_1 = \hat{K}_1^{(t)}, \hat{K}_0 = \hat{K}_0^{(t)}$
	
\end{algorithm}

\section{Simulation Results}

\subsection{Simulation setup}
In this section, numerical simulations are conducted to evaluate the performance of the proposed algorithm. For the proposed detection scheme, CSCD-H0 is adopted under ${\cal H}_0$ and CSCD-H1 is adopted under ${\cal H}_1$, so the detector is named GLRT-CSCD for simplicity. Likewise, we have  GLRT-OMP algorithms for the detectors with   OMP-based estimator. 

Note that the angle estimation is crucial for the detection performance, we  compare the accuracy of different methods. The OMP-based angle estimators are compared with our proposed CSCD-H0 and CSCD-H1 algorithms in ${\cal H}_0$ and ${\cal H}_1$ scenarios, respectively. With the estimated angle, GLRT is applied to detect whether the first-order indirect path exists.

Other simulation parameters are set as follows:
\begin{enumerate}
	\item The radar operates at  79 GHz with carrier wavelength $\lambda = 3.8$mm.  The number of transmitting elements    $M_T = 6$  and receive element $M_R = 8$. We first
	take simulation in a uniform linear array (ULA) given by Fig.  \ref{fig:MIMO_array}a. Consider that  the sparse linear array (SLA) is widely used for  automotive radar, so we also verify the performance of an  SLA given  by Fig.  \ref{fig:MIMO_array}b.  
	\begin{figure*}[!h]
		\centering
		\subfloat[][]{\includegraphics[width=3 in]{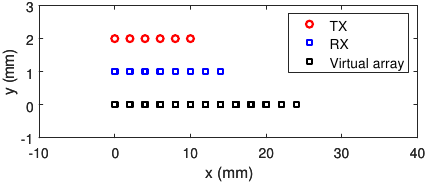}\label{fig:MIMO_ULA}}
		\subfloat[][]{\includegraphics[width=3 in]{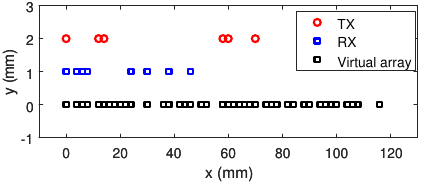}\label{fig:MIMO_SLA}}
		\caption{ Real and virtual layouts of the MIMO radar antennas, (a) ULA, (b) SLA.}
		\label{fig:MIMO_array}
	\end{figure*}
	\item The noise is randomly generated according to a Gaussian distribution with the variance $\sigma^2 = 1$. The path amplitudes  are generates according to
	$\bm{\beta} \sim  {\cal CN} (0, \sigma_{\beta}^2\bm{I}_{2K_1})$, $\bm{\alpha} \sim  {\cal CN} (0, \sigma_{\alpha}^2\bm{I}_{K_0})$. The  SNR of   direct paths and  first-order paths  are defined  as  $\rho_0= \sigma_{\alpha}^2/{\sigma^2}$ and  $\rho_1 = {\sigma_{\beta}^2}/{\sigma^2}$, respectively.
	\item 		The grids are  obtained by discretizing  angle  space $[-90^\circ,90^\circ]$ with a step of $2^\circ$.
	The max iteration of the  OMP, GOMP and CSCD-H0 and CSCD -H1 estimator are set to $T =10$. The stop criterion parameters are set as $I = 10, \epsilon  = \sqrt{\sigma^2M_TM_R}$, $\epsilon_1 = 0.4$ and $\epsilon_2 = 0$. For CSCD-H1, we set parameters   $\delta_r =  1$ and $J =  3$. 
	
	\item  We evaluate the  root-mean-squared-error (RMSE) of the angle estimation for the proposed algorithms. Notice that the algorithms return a bunch of estimations,  corresponding to either true paths or erroneous ones, and the  paths cannot be identified if there is no estimation  close to its direction.
	We thus refer to the RMSEs conditioned on the correct path estimation. In the undertaking simulation, a path is declared to be correctly estimated if its estimation error is smaller than the array beamwidth.
	Specifically, the  RMSEs 	of first-order path and direct path are calculated by
	\begin{eqnarray}
		\label{eq:rmse1}
		\operatorname{RMSE}_1=
		\sqrt{\splitfrac{\frac{1}{\mathrm{MC}} \sum\limits_{m=1}^{\mathrm{MC}}\frac{1}{2|\bm \Omega_1^m|} }{\cdot \sum\limits_{j \in \bm \Omega_1^m} \left(\splitfrac{({\vartheta}_j^{(m)}-\hat {\dot {\theta}}_j^{(m)})^2 }{+({\varphi}_j^{(m)}-\hat{{\varphi}}_j^{(m)})^2}\right)}},	
	\end{eqnarray}
	\begin{eqnarray}
		\label{eq:rmse0}
		\operatorname{RMSE}_0=\sqrt{\frac{1}{\mathrm{MC}} \sum_{m=1}^{\mathrm{MC}}\frac{1}{|\bm \Omega_0^m|} \sum_{j \in \bm \Omega_0^m} ({\theta}_j^{(m)}-\hat{\theta}_j^{(m)})^2},
	\end{eqnarray}
	respectively, where $\mathrm{MC}$ is the number of runs, $\bm \Omega_1^m$ and $\bm \Omega_0^m$ are the index set of the identified first-order paths and direct path in the $m$-th simulation  respectively; $|\cdot|$ denotes the cardinality of the input set; ${\vartheta}_j^{(m)}$, ${\varphi}_j^{(m)}$ are the  DOD and DOA the $j$-th first-order path in the $m$-th run and  ${\theta}_j^{(m)}$  is the DOA of   $j$-th  direct path, while $\hat{{\vartheta}}_j^{(m)}$, $\hat{{\varphi}}_j^{(m)}$ and $\hat{\theta}_j^{(m)}$ are the estimates, respectively. 
	\item 
	The detection performance of the proposed GLRT detector is compared with the performance bound derived in Sec. \ref{sec:perfor_the}. Specifically, the upper bound of $P_d$ is calculated by \eqref{eq:pd_the} under perfect angle estimation.
	\item 	
	Unless specifically stated, the probability of false alarm is set to be $10^{-3}$, and the numbers of independent trials used for simulating the probabilities of false alarm and detection are $100/P_{fa}$ and $10^4$, respectively. 
\end{enumerate}

\subsection{Estimation performance }

\begin{figure*} 
	\centering
	
	\subfloat[][]{\includegraphics[width=3  in]{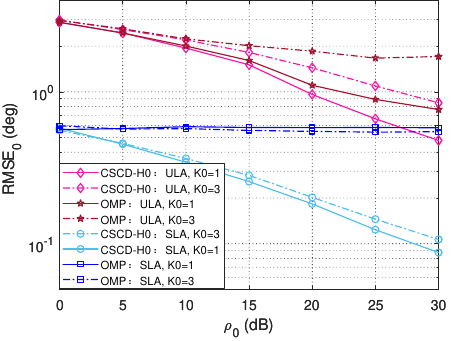}\label{fig:RMSE_H0}} 
	\subfloat[][]{\includegraphics[width=3  in]{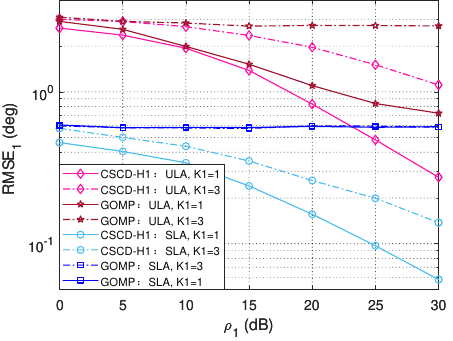}\label{fig:RMSE_H1}}
	\caption{ Plots of (a) $\operatorname{RMSE}_0$ in  ${\cal H }_0$ scenario, and (b) $\operatorname{RMSE}_1$ in  ${\cal H }_1$ scenario.}
	\label{fig:RMSE_dir_H0}
\end{figure*}

In this subsection, we verify the estimation performance of the proposed CSCD-H0 and CSCD-H1 algorithms. In Fig.  \ref{fig:RMSE_dir_H0}, we check the accuracy of direct path estimation in ${\cal H }_0$ scenario and first-order path estimation in ${\cal H }_1$ scenario.  As expected, the RMSE of proposed estimators decreases as $\rho_0$ or $\rho_1$ grows, indicating that larger SNR leads to better accuracy in estimation. OMP suffers from the off-grid issues, so its accuracy is consistently worse than that of the proposed algorithm. 

We notice that when the sparsity decreases ($K_0$ of Fig.  \ref{fig:RMSE_H0} or $K_1$ of Fig.  \ref{fig:RMSE_H1} increase from 1 to 3), a  decline in the accuracy could be observed. This phenomenon can be explained by many existing works in CS \cite{yang_sparsity}: the CS-based estimators take advantage of the sparsity inside signal for estimation  and the performance is getting worse as the sparsity decreases. 

\begin{figure*} 
	\centering
	\subfloat[][]{\includegraphics[width=3  in]{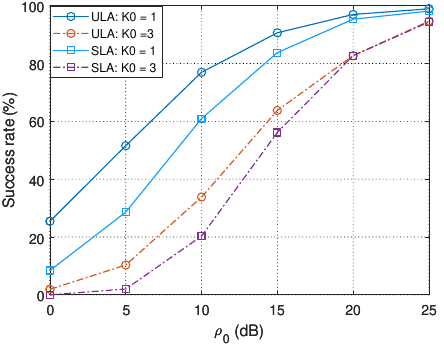}}
	\subfloat[][]{\includegraphics[width=3  in]{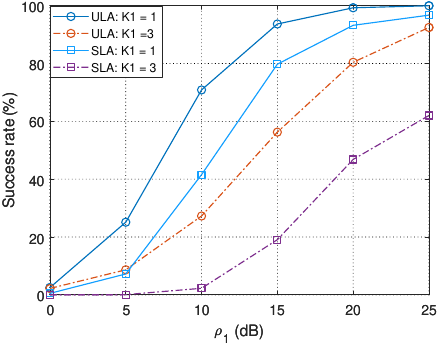}}
	\caption{ Success rate of (a) identifying all the direct paths in  ${\cal H }_0$ scenario and (b) identifying all the first-order paths in  ${\cal H }_1$ scenario}
	\label{fig:success_rate}
\end{figure*}

We also notice that the SLA, benefiting from its larger aperture, offers better accuracy compared to the ULA in both the direct paths estimation and the first-order paths estimation. This is partially due to the way we calculate the RMSE. While the accuracy is analyzed according to \eqref{eq:rmse1} and \eqref{eq:rmse0}, we only consider the paths that have been identified correctly, and the SLA usually provides better accuracy for the successfully identified paths.
However, as is  shown in Fig. \ref{fig:success_rate}, we observe that the SLA has a lower success rate of  identifying all paths  correctly than that of the ULA. This phenomenon is especially pronounced when $K_1 =3$,  the estimation method is more likely to make mistakes in identifying the path type, which could, in turn, reduce the performance of GLRT detection.

\begin{figure} 
	\centering
	\subfloat[][]{\includegraphics[width=1.6in]{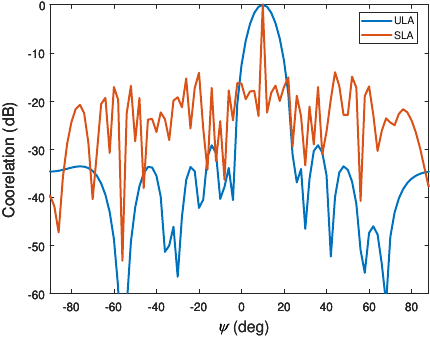}}
	\subfloat[][]{\includegraphics[width=1.6in]{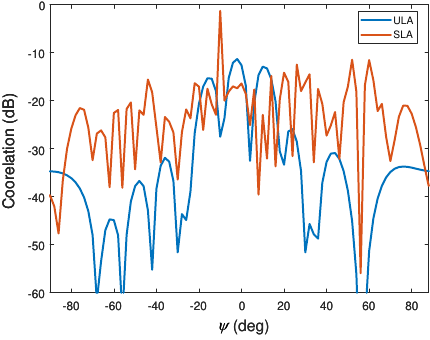}}
	\caption{ Comparison of (a) $\langle\mathbf{a}(\theta) ,\mathbf{a}({\psi})\rangle$ in ${\cal H}_0$ scenario, and (b) $ \langle\mathbf{a}_T(\vartheta)\circ\mathbf{a}_R(\varphi) ,\mathbf{a}({\psi})\rangle$ in ${\cal H}_1$ scenario.}
	\label{fig:correlation}
\end{figure}

{Unlike the ULA with half-lambda separation, the basis of SLA array could have large correlation. 
	In Fig.  \ref{fig:correlation}a, given a direct path $\theta =10^\circ$, we computed its correlation $\langle\mathbf{a}(\theta) ,\mathbf{a}({\psi})\rangle$  with the basis  ${\psi}\in [ -90^\circ,90^\circ] $ and observe that SLA has a narrower beamwidth but higher sidelobes.   Given a first-order path $(\vartheta, \varphi) = (10^\circ, -10^\circ)$, the correlation $ \langle\mathbf{a}_T(\vartheta)\circ\mathbf{a}_R(\varphi) ,\mathbf{a}({\psi})\rangle$ are plotted  in Fig.  \ref{fig:correlation}b, where a  distinct peak can be observed in SLA  even if the signals are not matched. It indicates that, 	in SLA,
	 the algorithm could make mistake when doing basis selection and the performance of GLRT  could be affected as well.
}

\subsection{Detection performance }
In order to assess the detection performance of the proposed system, we need at first to determine a method to set the detection threshold. In fact, unlike the ideal GLRT in \eqref{eq:GLRT_un}, the GLRT-CSCD detector   using CSCD-H0 and GCSD-H1 for estimation purposes  no longer exhibits CFAR behavior, due to the inevitable errors occurring in the estimation procedures outlined in the previous section. It is thus necessary at first to undertake a sensitivity analysis, in order to assess if outright adoption of the detection threshold of the ideal GLRT, as defined in \eqref{eq:Pfa}, yields a false alarm probability which at least preserves the order of magnitude of the designed value. To this end, we set a nominal value $P_{fa}=10^{-3}$, select the corresponding detection threshold through inversion of \eqref{eq:Pfa} and then evaluate the false alarm probability achieved by the GLRT-OMP and proposed GLRT-CSCD. The results are reported 
in Table \ref{Table:Pfa_simu}, assuming $\rho_0=\rho_1$. Even though our analysis is far from being exhaustive, the results clearly show that the actual false alarm probability of GLRT-CSCD stays below the nominal level for an ULA configuration under all the inspected values of $\rho_0$. The SLA configuration appears a little less favorable, especially as $K_0$ increases as analyzed in Fig.  \ref{fig:success_rate}. This is  due to the higher sidelobes that such an array configuration generates, with a consequent "spillover" of the direct paths into the first-order path subspace,  but the order of magnitude of the actual $P_{fa}$ is again preserved.  For the GLRT-OMP algorithm, the false alarm probability  for both ULA and SLA is higher. And in the SLA, the order of magnitude of the actual $P_{fa}$ can no longer be preserved. 
\begin{table*}[]
	\centering
		\renewcommand{\arraystretch}{1.3}
	\caption{Simulation of $P_{fa}$ with  $M_TM_R=48$ }\label{Table:Pfa_simu}
	\begin{tabular}{|c|c|ccc|ccc|}
		\hline
		\multirow{2}{*}{Array}  & \multirow{2}{*}{$K_0$} & \multicolumn{3}{c|}{GLRT-CSCD}                                                                                           & \multicolumn{3}{c|}{GLRT-OMP}                                                                                            \\ \cline{3-8} 
		&                     & \multicolumn{1}{c|}{$\rho_0=0$ dB} & \multicolumn{1}{c|}{$\rho_0=10$ dB} & $\rho_0=20$ dB & \multicolumn{1}{c|}{$\rho_0=0$ dB} & \multicolumn{1}{c|}{$\rho_0=10$ dB} & $\rho_0=20$ dB \\ \hline
		\multirow{3}{*}{ULA}    & 1                   & \multicolumn{1}{c|}{$1.74\times10^{-4}$}                         & \multicolumn{1}{c|}{$5.00\times10^{-5}$}                          &    $3.00\times10^{-5}$                     & \multicolumn{1}{c|}{$9.80\times10^{-3}$}                         & \multicolumn{1}{c|}{$4.30\times10^{-3}$}                          &   { $2.30\times10^{-3}$}                       \\ \cline{2-8} 
		& 2                   & \multicolumn{1}{c|}{$1.90\times10^{-4}$}                         & \multicolumn{1}{c|}{$8.00\times10^{-5}$}                          &   $2.00\times10^{-5}$                      & \multicolumn{1}{c|}{$7.70\times10^{-3}$}                         &  \multicolumn{1}{c|}{$1.60\times10^{-3}$}                          &  $7.00\times10^{-4}$                     \\ \cline{2-8} 
		& 3                   & \multicolumn{1}{c|}{$3.00\times10^{-4}$}                         & \multicolumn{1}{c|}{$1.90\times10^{-4}$}                          &       0                 & \multicolumn{1}{c|}{$5.80\times10^{-3}$}                         & \multicolumn{1}{c|}{$6.00\times10^{-4}$}                          &    $7.00\times10^{-4}$                    \\ \hline
		\multirow{3}{*}{SLA} & 1                   & \multicolumn{1}{c|}{$4.50\times10^{-4}$}                         & \multicolumn{1}{c|}{$5.00\times10^{-4}$}                          &    $1.50\times10^{-4}$                    & \multicolumn{1}{c|}{$2.12\times10^{-2}$}                         & \multicolumn{1}{c|}{$1.40\times10^{-2}$}                          &      $2.02\times10^{-2}$                   \\ \cline{2-8} 
		& 2                   & \multicolumn{1}{c|}{$5.80\times10^{-4}$}                         & \multicolumn{1}{c|}{$1.34\times10^{-3}$}                          &   $6.00\times10^{-4}$                      & \multicolumn{1}{c|}{$1.89\times10^{-2}$}                         & \multicolumn{1}{c|}{$8.30\times10^{-3}$}                          &        $7.90\times10^{-3}$                 \\ \cline{2-8} 
		& 3                   & \multicolumn{1}{c|}{$8.50\times10^{-4}$}                         & \multicolumn{1}{c|}{$3.68\times10^{-3}$}                          &      $5.30\times10^{-4}$                 & \multicolumn{1}{c|}{$1.68\times10^{-2}$}                         & \multicolumn{1}{c|}{$5.70\times10^{-3}$}                          &$3.20\times10^{-3}$                        \\ \hline
	\end{tabular}
\end{table*}

\begin{figure*} 
	\centering
	\subfloat[][]{\includegraphics[width=3 in]{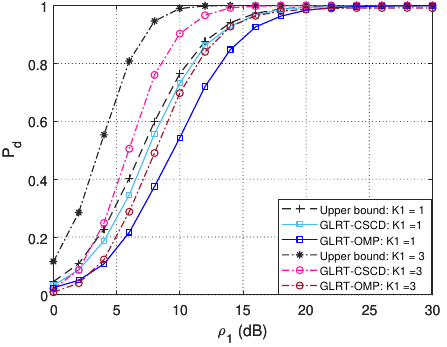}\label{fig:Pd_difK_ULA}}
	\subfloat[][]{\includegraphics[width=3 in]{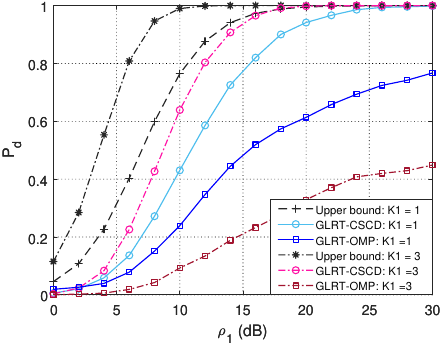}\label{fig:Pd_difK_SLA}}
	\caption{ $P_d$ versus $\rho_1$ for (a) ULA, and (b) SLA.}
	\label{fig:Pd_difK}
\end{figure*}

\begin{figure}
	\centering
	\includegraphics[width=3in]{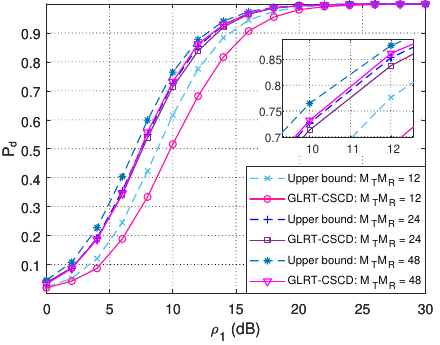}
	\caption{   $P_d$ versus $\rho_1$ for different $M_TM_R$.	}
	\label{fig:Pd_difMtMr}
\end{figure} 
In Fig.  \ref{fig:Pd_difK},  the $P_d$  of GLRT-OMP, and GLRT-CSCD are compared with the upper bound.   For the ULA results given by Fig. \ref{fig:Pd_difK}a, the detection performance of GLRT-CSCD with $K_1 = 1$ is  close to the upper bound. As $K_1 =3$,  the performance gap between the proposed detectors and the upper bound becomes larger due to the degradation in estimation performance.  This is also validated by our RMSE simulation given by Fig. \ref{fig:RMSE_H1}.
In the SLA,  we can see from Fig. \ref{fig:Pd_difK}b, the  discrepancy between the proposed detectors and the upper bound is larger than that of ULA. The proposed GLRT-CSCD still benefits from a larger $K_1$ to achieve better detection performance. However, the angle estimation performance of the OMP is much worse than that of the proposed algorithm, so its detection performance is considerably below the upper bound. 

To compare detection performance across different array sizes, we set up  simulation with $M_T=3$, $M_R=4$ ($M_TM_R=12$), $M_T=4$, $M_R=6$ ($M_TM_R=24$) and $M_T=6$, $M_R=8$ ($M_TM_R=48$).  For simplicity, we adopt ULAs with half-wavelength element spacing and the detection performance are evaluated when  when $K_0=1$ and $K_1 =1$.
As reported in Fig. \ref{fig:Pd_difMtMr},  the simulated performance is close to the upper bound given by the theoretical analysis. As indicated by both results, the detection performance becomes better as the array sizes $M_TM_R$ grows.

\section{Conclusions}
In this paper, we investigate the identification of ghost targets for automotive radar in the presence of multipaths. The existence of indirect paths is modeled as a binary composite hypothesis test and a GLRT detector is proposed to determine whether indirect paths exist in a delay-Doppler cell. If a cell contains indirect paths, the ghost targets  could be removed and the desired direct paths can be preserved. To estimate the angle of both direct and indirect paths, we propose a  sparsity-enforced CS approach to estimate the angular parameters in the continuous domain. Simulation results indicate that the proposed algorithm outperforms on-grid CS estimators, thereby leading to better detection performance. The theoretical detection performance of the proposed GLRT has  been analyzed under perfect angle estimation. Simulation shows that the false alarm rate of the proposed detector could be controlled and the detection performance is close to the theoretical bound in ULA case. 

\appendix
\subsection{Derivation of   $\mathbf{g}_0$ in \eqref{eq:g0_h0} and $\mathbf{H}_0$ in \eqref{eq:H0_h0} }
For clarity, we drop the superscript $(t,i)$ and input variable of the functions in some of the following derivation, i.e. $ F = F(\widehat{\bm{\Theta}}_0^{(t,i)} )$ and $ \mathbf{A}_0 = \mathbf{A}(\widehat{\bm{\Theta}}^{(t,i)}_0)$.

Denote $F =  \mathbf{f}^H \mathbf{f}$ with $\mathbf{f} = \bm z - \mathbf{A}\mathbf{A}^\dagger \bm z $, 
the gradient of $F$ with respect to $\bm{\Theta}_0 \in \mathbb{R}^{ {K}_0\times 1}$ can be calculated by
\begin{eqnarray}	\mathbf{g}_0 &=& \left[\frac{\partial{F}}{\partial{  \theta_{1}}},\frac{\partial{F}}{\partial{  \theta_{2}}},\dots,\frac{\partial{F}}{\partial{  \theta_{{K}_0}}}\right]^T,
\end{eqnarray}
where the    $q$-th element $[\mathbf{g}_0]_q$ given as $	\frac{\partial{{F}}}{\partial{  \theta_{q}}} =2{\rm{Re}}((\frac{\partial \mathbf{f}}{\partial{   \theta_{q}}})^H \mathbf{f})$. 
Following the derivation in    \cite{Ottersten1993ExactAL}, we  obtain 
\begin{eqnarray}
	\label{eq:g0_q}
	\begin{aligned}
		[\mathbf{g}_0]_q = -2{\rm{Re}}\left\{{\rm Tr}\{\mathbf{A}_0^\dagger  \bm z\bm z^H  \bm{P}_0\mathbf{A}_{q}\}\right\},
	\end{aligned}
\end{eqnarray}
where 	$\mathbf{A}_{q } = \frac{\partial{\mathbf{A}_0 }}{\partial{ \theta}_{q}} = [\mathbf{0},\mathbf{0},\dots,\frac{\partial {\mathbf{a}}}{\partial{ \theta_{q}}},\dots,\mathbf{0}]$ with $\frac{\partial {\mathbf{a}}}{\partial{ \theta_{q}}} = \frac{\partial{{\mathbf{a}}( \theta_{q}) }} {\partial{ \theta_{q}}}$.

The Hessian $\mathbf{H}_{0}$ denotes approximate  second order partial derivative of $F$ with respect to  $\bm{\Theta}_0$.  In this matrix,  the $(q,p)$-th element is denoted as $	[\mathbf{H}_0]_{q,p} =   2{\rm{Re}}\left\{ \left(\frac{\partial \mathbf{f} }{\partial  \theta_{q}}\right)^H \frac{\partial \mathbf{f} }{\partial  \theta_{p}}\right\}$  and can be calculated as  follows
\begin{eqnarray}
	\label{eq:Hqp}
	\begin{aligned}
		[\mathbf{H}_0]_{q,p}
		= &2{\rm{Re}}\left\{{\rm Tr}\{\mathbf{A}_{p}\mathbf{A}_0^\dagger\bm z \bm z^H(\mathbf{A}_0^\dagger)^H  \mathbf{A}_{q}^H
		\bm{P}_0\}\right\} \\
		&+ 2{\rm{Re}}\left\{{\rm Tr}\{\mathbf{A}_{p}^H\bm{P}_0\bm z\bm z^H  \bm{P}_0\mathbf{A}_{q}\mathbf{A}_0^\dagger (\mathbf{A}_0^\dagger)^H\}\right\}.
	\end{aligned}
\end{eqnarray}

Defining a partial matrix $\mathbf{D}_{0} = \left[\frac{\partial {\mathbf{a}}}{\partial  	\theta_1},\frac{\partial {\mathbf{a}}}{\partial  \theta_{2}},\dots,\frac{\partial {\mathbf{a}}}{\partial \theta_{K_0}}\right]$,  then  the  matrix form of $\mathbf{g} $ and    $\mathbf{H}_0$ can be given by \eqref{eq:g0_h0} and \eqref{eq:H0_h0}, respectively.
\color{black}
\subsection{Derivation of  $\mathbf{g}$'s in \eqref{eq:g_group} and $\mathbf{H}$'s in \eqref{eq:H_group}}
For clarity, we drop the  superscript and input variable of the functions in some of the following derivation, i.e. $ \bar F = \bar F(\bar{\bm{\Theta}}^{(t,i)} )$ and $ \mathbf{A} = \mathbf{A}(\bar{\bm{\Theta}}^{(t,i)},\bar{\bm{\Phi}}^{(t,i)})$.
 In the following, we derive the matrix expression of $\mathbf g _T$ and $\mathbf H _{\rm TT}$, the derivation  for $\mathbf g _{\rm R}$, $\mathbf g' _0$,  $\mathbf H _{\rm TR}$, $\mathbf H _{\rm RR}$, $\mathbf H _{\rm RT}$, $\mathbf H _{\rm 0T}$, $\mathbf H _{\rm T0}$, $\mathbf H _{\rm R0}$, $\mathbf H _{\rm 0R}$,  $\mathbf H _{00}$ follow  similar arguments and are omitted for brevity.
\label{gradient_hessian_deriv}
Similar with \eqref{eq:g0_q}, we know the $q$-th element of $\mathbf g _T$ can be given as
\begin{eqnarray}
	\label{eq:g_T_q}
		[\mathbf{g}_{\rm T}]_q 
		&=&	 -2{\rm{Re}}\{{\rm Tr}\{\mathbf{A}^\dagger  \bm z\bm z^H  \bm{P}_1\mathbf{A}'_{q}\}\},\nonumber\\
	&=&-2{\rm{Re}}\left\{{\rm Tr}\{\bm{\Gamma}\mathbf{A}'_q\}\right\},
\end{eqnarray}
where $\bm{\Gamma} = \mathbf{A}^\dagger \bm z\bm z^H \bm{P}_1\in \mathbb{C}^{(2K_1+K_0)\times M_TM_R}$, $\mathbf{A}'_{q} = \frac{\partial{\mathbf{A} }}{\partial{ \vartheta}_{q}} = [\mathbf{0},\mathbf{0},\dots,\frac{\partial {\mathbf{a}_1}}{\partial{ \vartheta_{q}}},\dots,\mathbf{0},\dots,\frac{\partial {\mathbf{a}_2}}{\partial{ \vartheta_{q}}},\dots,\mathbf{0}]$ with $\frac{\partial {\mathbf{a}_1}}{\partial{ \vartheta_{q}}} = \frac{\partial {\mathbf{a}_T(\vartheta_q)\otimes\mathbf{a}_R(\varphi_q)}}{\partial{ \vartheta_{q}}}$ and $\frac{\partial {\mathbf{a}_2}}{\partial{ \vartheta_{q}}} = \frac{\partial {\mathbf{a}_T(\varphi_q)\otimes\mathbf{a}_R(\vartheta_q)}}{\partial{ \vartheta_{q}}}$. 
We divide the matrix  $\bm{\Gamma}$ into three submatrices, denoted  as $\bm{\Gamma} = [\bm{\Gamma}_1, \bm{\Gamma}_2, \bm{\Gamma}_0]$, where $\bm{\Gamma}_1, \bm{\Gamma}_2 \in \mathbb{C}^{K_1\times M_TM_R}$, $\bm{\Gamma}_0 \in \mathbb{C}^{K_0\times M_TM_R} $. Then  	
 \eqref{eq:g_T_q} can be rewritten as 
\begin{eqnarray}
	[\mathbf{g}_{\rm T}]_q = -2{\rm{Re}}\left\{\bm{\Gamma}^T_1(q)(\frac{\partial{\mathbf{a}_1}}{\partial{\vartheta_q}})^T+\bm{\Gamma}^T_2(q)(\frac{\partial{\mathbf{a}_2}}{\partial{\vartheta_q}})^T\right\},
\end{eqnarray}
where $\bm \Gamma_1^T(q)$ and $\bm \Gamma_2^T(q)$ denote the $q$ row of $\bm{\Gamma}_1$ and $\bm{\Gamma}_2$, respectively.
Define  two partial matrices: $\mathbf{D}_{\rm T1} = \left[\frac{\partial \mathbf{a}_1}{\partial \vartheta_{1}},\frac{\partial \mathbf{a}_1}{\partial \vartheta_{2}},\dots,\frac{\partial \mathbf{a}_1}{\partial \vartheta_{K_1}}\right]$, $	\mathbf{D}_{\rm T2} = \left[\frac{\partial \mathbf{a}_2}{\partial \vartheta_{2}},\frac{\partial \mathbf{a}_2}{\partial\vartheta_{2}},\dots,\frac{\partial \mathbf{a}_2}{\partial\vartheta_{K_1}}\right]$.
We can then obtain the matrix form of  $\mathbf{g}_{\rm T}$ given by 
\begin{eqnarray}
	\label{eq:g_DOD}
	\mathbf{g}_{\rm T} 
	&=&-2{\rm{Re}}\{\rm{diag}\{\bm{\Gamma}_1 \mathbf{D}_{\rm T1}+\bm{\Gamma}_2 \mathbf{D}_{\rm T2}\}\}.
\end{eqnarray} 
Similarly, we define $	\mathbf{D}_{\rm R1} = \left[\frac{\partial \mathbf{a}_1}{\partial \varphi_{1}},\frac{\partial \mathbf{a}_1}{\partial \varphi_{2}},\dots,\frac{\partial \mathbf{a}_1}{\partial \varphi_{K_1}}\right]$, $		\mathbf{D}_{\rm R2} = \left[\frac{\partial \mathbf{a}_2}{\partial \varphi_{2}},\frac{\partial \mathbf{a}_2}{\partial\varphi_{2}},\dots,\frac{\partial \mathbf{a}_2}{\partial\varphi_{K_1}}\right]$ and $			\mathbf{D}_{0} = \left[\frac{\partial {\mathbf{a}}}{\partial  	\theta_1},\frac{\partial {\mathbf{a}}}{\partial  \theta_{2}},\dots,\frac{\partial {\mathbf{a}}}{\partial \theta_{K_0}}\right]$, and  obtain  
\begin{eqnarray}
	\label{eq:g_DOA}
	\mathbf{g}_{\rm R} 	&=&-2{\rm{Re}}\{\rm{diag}\{{\bm{\Gamma}}_1 \mathbf{D}_{\rm R1}+{\bm{\Gamma}}_2 \mathbf{D}_{\rm R2}\}\},\\
	\label{eq:g_dir}
	\mathbf{g}'_0 &=&-2{\rm{Re}}\{\rm{diag}\{\bm{\Gamma}_0 \mathbf{D}_0\}\}.
\end{eqnarray}

The Hessian $\mathbf{H}_{\rm TT}$ denotes second order partial derivative with respect to  $\bm{\Theta}_1$, in which the $(q,p)$-th element  is 
\begin{eqnarray}
	\label{eq:Hqp_h1}
	\begin{aligned}
	[\mathbf{H}_{\rm TT}]_{q,p} 			
		= &2{\rm{Re}}\left\{{\rm Tr}\{\mathbf{A}'_{p}\mathbf{A}^\dagger\bm z \bm z^H(\mathbf{A}^\dagger)^H  {\mathbf{A}'_{q}}^H
		\bm{P}_1\}\right\} \\
		&+ 2{\rm{Re}}\left\{{\rm Tr}\{{\mathbf{A}'_{p}}^H\bm{P}_1\bm z\bm z^H  \bm{P}_1\mathbf{A}'_{q}\mathbf{A}^\dagger (\mathbf{A}^\dagger)^H\}\right\},
	\end{aligned}
\end{eqnarray}
 where the first item  
\begin{eqnarray}
	\begin{aligned}
		{\rm Tr}\{&\mathbf{A}'_{p}\mathbf{A}^\dagger\bm z \bm z^H(\mathbf{A}^\dagger)^H  {\mathbf{A}'_{q}}^H\bm{P}_1\}\nonumber\\
		=&{\mathbf{S}}_{p,q}(\frac{\partial \mathbf{a}_1}{\partial\vartheta_{q}})^H \bm{P}_1\frac{\partial \mathbf{a}_1}{\partial\vartheta_{p}} +[\mathbf{S}]_{p,q+K_1}(\frac{\partial \mathbf{a}_2}{\partial\vartheta_{q}})^H\bm{P}_1\frac{\partial \mathbf{a}_1}{\partial\vartheta_{p}}\nonumber	\\
		+&[\mathbf{S}]_{p+K_1,q}(\frac{\partial \mathbf{a}_1}{\partial\vartheta_{q}})^H\bm{P}_1\frac{\partial \mathbf{a}_2}{\partial\vartheta_{p}}+[\mathbf{S}]_{p+K_1,q+K_1}(\frac{\partial \mathbf{a}_2}{\partial\vartheta_{q}})^H \bm{P}_1\frac{\partial \mathbf{a}_2}{\partial\vartheta_{p}}
	\end{aligned}
\end{eqnarray}
with $	\mathbf{S} = \mathbf{A}^\dagger\bm z \bm z^H(\mathbf{A}^\dagger)^H$, 
and  the second item can be rewritten as
\begin{eqnarray}
	\begin{aligned}
		{\rm Tr}\{&{\mathbf{A}'_{p}}^H\bm{P}_1 \bm z	\bm z^H  \bm{P}_1\mathbf{A}'_{q}\mathbf{A}^\dagger (\mathbf{A}^\dagger)^H \}\nonumber\\
		= 	&[\mathbf{C}]_{q,p} (\frac{\partial \mathbf{a}_1}{\partial\vartheta_{p}})^H \mathbf{X}\frac{\partial \mathbf{a}_1}{\partial\vartheta_{q}} + [\mathbf{C}]_{q+K_1,p}(\frac{\partial \mathbf{a}_1}{\partial\vartheta_{p}})^H \mathbf{X}\frac{\partial \mathbf{a}_2}{\partial\vartheta_{q}}\nonumber\\
		 + &[\mathbf{C}]_{q,p+K_1}(\frac{\partial \mathbf{a}_2}{\partial\vartheta_{p}})^H \mathbf{X}\frac{\partial \mathbf{a}_1}{\partial\vartheta_{q}}+ [\mathbf{C}]_{q+K_1,p+K_1}(\frac{\partial \mathbf{a}_2}{\partial\vartheta_{p}})^H \mathbf{X}\frac{\partial \mathbf{a}_2}{\partial\vartheta_{q}}
	\end{aligned}
\end{eqnarray}
with $\mathbf{X} =  \bm{P}_1\bm z\bm z^H  \bm{P}_1$ and $\mathbf{C} = \mathbf{A}^\dagger(\mathbf{A}^\dagger)^H$.
		To represent $\mathbf{H}_{\rm TT}$ in matrix form, we divide matrices $\mathbf{S}$ and $\mathbf{C}$ into 
		\begin{eqnarray}
				\mathbf{S} = \left[ \begin{array}{cc}
						\mathbf{S}_1 & \mathbf{S}_{10} \\
						\mathbf{S}_{01} &  \mathbf{S}_{0} 
					\end{array} 
				\right],\\
				\mathbf{C} = \left[ \begin{array}{cc}
						\mathbf{C}_1 & \mathbf{C}_{10} \\
						\mathbf{C}_{01} &  \mathbf{C}_{0} 
					\end{array} 
				\right],
			\end{eqnarray}
		where $\mathbf{S}_1, \mathbf{C}_1 \in \mathbb{C}^{2K_1\times2K_1}$,$\mathbf{S}_{10}, \mathbf{C}_{10}  \in \mathbb{C}^{2K_1\times K_0}$, $\mathbf{S}_{01}, \mathbf{C}_{01}  \in \mathbb{C}^{K_0\times 2K_1}$ and  $\mathbf{S}_0, \mathbf{C}_0 \in \mathbb{C}^{K_0\times K_0}$.
		Then, we  obtain
		\begin{eqnarray}
			\label{eq:HTT}
				\mathbf{H}_{\rm TT}&=&2{\rm{Re}}\left\{ \mathbf{E}_{\rm h} (\mathbf{D}_{\rm T})^H \bm{P}_1\mathbf{D}_{\rm T} \odot \mathbf{S}_1^T \mathbf{E}_{\rm h}^T\right\}\nonumber\\
				&&+2{\rm{Re}}\left\{ \mathbf{E}_{\rm h} ( (\mathbf{D}_{\rm T})^H \mathbf{X} \mathbf{D}_{\rm T})^T\odot\mathbf{C}_1 \mathbf{E}_{\rm h}^T\right\},
			\end{eqnarray}
		where $\mathbf{E}_{\rm h}  = [\bm{I}_{K_1}, \bm{I}_{K_1}] \in \mathbb{R}^{{K_1}\times 2{K_1}}$, $\mathbf{D}_{\rm T}  =[\mathbf{D}_{\rm T1}, \mathbf{D}_{\rm T2}]$.
		Similarly, we define  $\mathbf{D}_{\rm R}  =[\mathbf{D}_{\rm R1}, \mathbf{D}_{\rm R2}]$ and obtain 
		\begin{eqnarray}
			\label{eq:HTR}
				\mathbf{H}_{\rm TR}&=&2{\rm{Re}}\left\{ \mathbf{E}_{\rm h} (\mathbf{D}_{\rm T})^H \bm{P}_1\mathbf{D}_{\rm R} \odot \mathbf{S}^T \mathbf{E}_{\rm h}^T\right\}\nonumber\\
				&&+2{\rm{Re}}\left\{ \mathbf{E}_{\rm h} ( (\mathbf{D}_{\rm R})^H \mathbf{X} \mathbf{D}_{\rm T})^T\odot\mathbf{C} \mathbf{E}_{\rm h}^T\right\},
			\end{eqnarray}  
		\begin{eqnarray}
			\label{eq:HRT}
				\mathbf{H}_{\rm RT}&=&2{\rm{Re}}\left\{ \mathbf{E}_{\rm h} (\mathbf{D}_{\rm R})^H \bm{P}_1\mathbf{D}_{\rm T} \odot \mathbf{S}_1^T \mathbf{E}_{\rm h}^T\right\}\nonumber\\
				&&	+2{\rm{Re}}\left\{ \mathbf{E}_{\rm h} ( (\mathbf{D}_{\rm T})^H \mathbf{X} \mathbf{D}_{\rm R})^T\odot\mathbf{C}_1 \mathbf{E}_{\rm h}^T\right\},
			\end{eqnarray}
		\begin{eqnarray}
			\label{eq:HRR}
				\mathbf{H}_{\rm RR}&=&2{\rm{Re}}\left\{ \mathbf{E}_{\rm h} (\mathbf{D}_{\rm R})^H \bm{P}_1\mathbf{D}_{\rm R} \odot \mathbf{S}_1^T \mathbf{E}_{\rm h}^T\right\}\nonumber\\
				&&	+2{\rm{Re}}\left\{ \mathbf{E}_{\rm h} ( (\mathbf{D}_{\rm R})^H \mathbf{X} \mathbf{D}_{\rm R})^T\odot\mathbf{C}_1 \mathbf{E}_{\rm h}^T\right\},
			\end{eqnarray}
		\begin{eqnarray}
			\label{eq:HT0}
				\mathbf{H}_{\rm T0}&=&2{\rm{Re}}\left\{ \mathbf{E}_{\rm h}  \left((\mathbf{D}_{\rm T})^H \bm{P}_1\mathbf{D}_0 \odot \mathbf{S}_{01}^T\right)\right\}\nonumber\\
				&&+2{\rm{Re}}\left\{ \mathbf{E}_{\rm h} \left(( (\mathbf{D}_{0})^H \mathbf{X} \mathbf{D}_T)^T\odot\mathbf{C}_{10}\right)\right\},
			\end{eqnarray}
		\begin{eqnarray}
			\label{eq:HR0}
				\mathbf{H}_{\rm R0}&=&2{\rm{Re}}\left\{ \mathbf{E}_{\rm h}  \left((\mathbf{D}_{\rm R})^H \bm{P}_1\mathbf{D}_0 \odot \mathbf{S}_{01}^T\right)\right\}\nonumber\\
				&&+2{\rm{Re}}\left\{ \mathbf{E}_{\rm h} \left(( (\mathbf{D}_{0})^H \mathbf{X} \mathbf{D}_R)^T\odot\mathbf{C}_{10}\right)\right\},
			\end{eqnarray}
		\begin{eqnarray}
			\label{eq:H0T}
				\mathbf{H}_{\rm 0T}&=&2{\rm{Re}}\left\{   \left((\mathbf{D}_0)^H \bm{P}_1\mathbf{D}_{\rm T} \odot \mathbf{S}_{10}^T\right)\mathbf{E}_{\rm h}^T\right\}\nonumber\\
				&&+2{\rm{Re}}\left\{  \left(( (\mathbf{D}_{\rm T})^H \mathbf{X} \mathbf{D}_{0})^T\odot\mathbf{C}_{01}\right)\mathbf{E}_{\rm h}^T\right\},
			\end{eqnarray}
		\begin{eqnarray}
			\label{eq:H0R}
				\mathbf{H}_{\rm 0R}&=&2{\rm{Re}}\left\{   \left((\mathbf{D}_0)^H \bm{P}_1\mathbf{D}_{\rm R} \odot \mathbf{S}_{10}^T\right)\mathbf{E}_{\rm h}^T\right\}\nonumber\\
				&&+2{\rm{Re}}\left\{  \left(( (\mathbf{D}_{\rm R})^H \mathbf{X} \mathbf{D}_{0})^T\odot\mathbf{C}_{01}\right)\mathbf{E}_{\rm h}^T\right\},
			\end{eqnarray}
		\begin{eqnarray}
			\label{eq:H00}
				\mathbf{H}_{00}&=&2{\rm{Re}}\left\{   \mathbf{D}_0^H \bm{P}_1\mathbf{D}_{0} \odot \mathbf{S}_{0}^T\right\}\nonumber\\
				&&+2{\rm{Re}}\left\{ ( \mathbf{D}_0^H \mathbf{X} \mathbf{D}_{0})^T\odot\mathbf{C}_0\right\}.
			\end{eqnarray}
\bibliographystyle{IEEEtran}
\bibliography{database}

\end{document}